\documentclass[journal]{IEEEtran}

\usepackage{booktabs}
\usepackage{threeparttable}
\usepackage{multirow}
\usepackage[table,xcdraw]{xcolor}
\usepackage{bm}
\usepackage{amsthm}
\usepackage{fontawesome5}
\usepackage {tcolorbox}
\usepackage{arydshln}
\usepackage{hyperref} 
\usepackage{amsmath,amsfonts}
\usepackage{algorithmic}
\usepackage{array}
\usepackage[caption=false,font=normalsize,labelfont=sf,textfont=sf]{subfig}
\usepackage{textcomp}
\usepackage{stfloats}
\usepackage{url}
\usepackage{verbatim}
\usepackage{graphicx}
\usepackage{amssymb}
\usepackage[ruled,linesnumbered]{algorithm2e}
\def\BibTeX{{\rm B\kern-.05em{\sc i\kern-.025em b}\kern-.08em
    T\kern-.1667em\lower.7ex\hbox{E}\kern-.125emX}}
\usepackage{balance}
\usepackage{cite}
\usepackage[dvipsnames]{xcolor}

\definecolor{highlight1}{HTML}{C0C0C0}
\definecolor{highlight2}{HTML}{DCDCDC}
\definecolor{highlight3}{HTML}{F5F5F5}

\newcommand{\citet}[1]{\cite{#1}}

\newcommand{\rqbox}[1]{
  \noindent\faIcon{question-circle} \textbf{RQ\arabic{rqcounter}: #1} \\
  \stepcounter{rqcounter}
}
\newcounter{rqcounter}
\newtcolorbox{boxK}{
    top=2pt,
    bottom=2pt,
    left=2pt,
    right=2pt,
    boxrule = 0pt,
    toprule = 0pt, 
}

\newcommand{\findingbox}[1]{
    \begin{boxK} 
        \small \faIcon{lightbulb} \textbf{Finding \arabic{findingcounter}:} #1
    \end{boxK}
    \stepcounter{findingcounter}
}
\newcounter{findingcounter}
\newtheorem{theorem}{Theorem}
\newtheorem{definition}[theorem]{Definition}
\newtheorem{lemma}[theorem]{Lemma}
\newtheorem{proposition}[theorem]{Proposition}

\newcommand{\trans}{\mathcal{P}_{\text{trans}}}
\newcommand{\fp}{\mathrm{fp}}

\usepackage[capitalize]{cleveref}
\crefname{section}{Sec.}{Secs.}
\Crefname{section}{Section}{Sections}
\Crefname{table}{Table}{Tables}
\crefname{table}{Tab.}{Tabs.}

\usepackage{colortbl} 
\usepackage{xcolor}   
\usepackage{booktabs} 
\usepackage{pgfmath}  
\usepackage{listings}

\definecolor{blue}{HTML}{000000}
\definecolor{dgreen}{RGB}{0, 0, 0}
\lstdefinestyle{customcpp}{
    language=C++,
    basicstyle=\ttfamily\small,
    keywordstyle=\color{Maroon},        
    stringstyle=\color{DarkOrchid},     
    commentstyle=\color{ForestGreen},   
    moredelim=**[is][\color{DarkGoldenrod}]{@}{@}, 
}

\definecolor{low}{HTML}{1B5E20} 
\definecolor{high}{HTML}{E8F5E9} 

\newcommand{\minvalue}{63.0}
\newcommand{\maxvalue}{66.0}
\newcommand{\gradientcell}[1]{%
    \pgfmathsetmacro{\percentage}{(100.0 * ((#1-\minvalue)/(\maxvalue-\minvalue))^0.5 )}
    \edef\calculatedcolor{\noexpand\cellcolor{low!\percentage!high}}
    \calculatedcolor #1\%
}

\begin{document}

\title{Post-Incorporating Code Structural Knowledge into Pretrained Models via ICL for Code Translation}

\author{
	Yali~Du$^\dagger$,
    Hui~Sun$^\dagger$,
    and~Ming~Li,~\IEEEmembership{Member,~IEEE},
\IEEEcompsocitemizethanks{
\IEEEcompsocthanksitem 
The authors are with the National Key Laboratory for Novel Software Technology, Nanjing University, Nanjing 210023, China, and also with the School of Artificial Intelligence, Nanjing University, Nanjing 210023, China. \protect\\
Email: \{duyl, sunh, lim\}@lamda.nju.edu.cn%
\IEEEcompsocthanksitem 
$^\dagger$ These authors contributed equally.
Corresponding author: Ming Li.
}
\thanks{Manuscript received August 27, 2025.}}



\maketitle

\begin{abstract}
Code translation migrates codebases across programming languages. 
Recently, large language models (LLMs) have achieved significant advancements in software mining.
However, handling the syntactic structure of source code remains a challenge.
Classic syntax-aware methods depend on intricate model architectures and loss functions, 
rendering their integration into LLM training resource-intensive.
This paper employs in-context learning (ICL), which directly integrates task exemplars into the input context, to post-incorporate code structural knowledge into pre-trained LLMs.
We revisit exemplar selection in ICL from an information-theoretic perspective, 
proposing that list-wise selection based on information coverage is more precise and general objective than traditional methods based on combine similarity and diversity.
To address the challenges of quantifying information coverage, 
we introduce a surrogate measure, Coverage of Abstract Syntax Tree~(CAST), 
measuring maximum subtree coverage between ASTs of test source code and exemplars.
Furthermore, we formulate the NP-hard CAST maximization for exemplar selection and prove that it is a standard submodular maximization problem.
Therefore, we propose a greedy algorithm for CAST submodular maximization, which theoretically guarantees a $(1-1/e)$-approximate solution in polynomial time complexity.
Our method is the first training-free and model-agnostic approach to post-incorporate code structural knowledge into existing LLMs at test time.
Experimental results show that our method significantly improves LLMs performance in code translation and reveals two meaningful insights:
1)~Code structural knowledge can be effectively post-incorporated into pre-trained LLMs during inference, despite being overlooked during training; 
2)~Scaling up model size or training data does not lead to the emergence of code structural knowledge, underscoring the necessity of explicitly considering code syntactic structure.
\end{abstract}

\begin{IEEEkeywords}
    Code Translation, Syntactic Validity, Large Language Model, In-context Learning.
\end{IEEEkeywords}

\section{Introduction}
\IEEEPARstart{C}{ode} translation migrates codebases from one programming language to another,  
avoiding the time- and labor-intensive process of redeveloping software from scratch.
Automated code translation significantly enhances coding productivity in various practical scenarios, such as: 
1)~modernizing legacy systems~(e.g., COBOL or FORTRAN) with contemporary languages~(e.g., Python or Java) to extend functionality;
2)~ensuring cross-platform compatibility or using language-specific advantages, e.g., integrating high-performance C++ code with Python-based machine learning frameworks.

Recently, large language models~(LLMs) have demonstrated exceptional performance across various software mining tasks~\cite{guo2024deepseek,nijkamp2022codegen,codellama,li2023starcoder,liurb2024icml,lixy2025arxiv,pan2024lost}. 
However, effectively handling the syntactic structure of source code remains a significant challenge.
Unlike the grammatical structure of natural language, the syntactic structure of source code is stricter and more complex, as it directly reflects program behavior. 
Previous studies have shown that explicitly incorporating code syntactic knowledge can significantly enhance performance in software mining tasks~\cite{treebert,tree2tree,code2vec,sgattention,semanticcodebert,GraphCodeBERT,atom,astt5}. 
However, traditional methods often depend on complex model architectures and customized loss functions, making their integration with LLMs both time-consuming and resource-intensive, even infeasible for proprietary models.

In-context learning~(ICL) provides an effective and efficient solution of integrating new domain-specific knowledge into off-the-shelf LLMs without additional training by directly incorporating task exemplars into the input context~\cite{luo2024context}. 
For instance, in Python-to-C++ code translation, 
ICL enhances LLMs performance by incorporating a few paired exemplars~(e.g., Python-to-C++ translation pairs) into the input prompt. 
This training-free approach offers flexibility across diverse tasks, avoids catastrophic forgetting, and ensures efficient resource utilization.
Selecting appropriate contextual exemplars is critical for ICL. 
However, research on exemplar selection strategies for ICL in code translation is still limited.

Traditional ICL exemplar selection strategies often overlook code translation tasks and fail to consider the critical syntactic information inherent in source code~\cite{luo2024context}.
These methods primarily rely on sample-wise similarity~\cite{liu2021makes,gao2022transform,luo-etal-2021-weakly}, ignoring compositional relationships among exemplars, which often leads to redundancy.
Furthermore, syntax-aware similarity remains underexplored.
Some approaches attempt to ensure diversity through clustering and additional regularization losses~\cite{li2024long,li2022self,li2023mot,ye2023compositional,levy2022diverse}.
However, these methods often hinder the optimization of relevance and overlook compositional relationships in the syntactic structure of code.
Moreover, balancing similarity and diversity poses a significant challenge.

\IEEEpubidadjcol

This paper revisits exemplar selection for ICL from an information-theoretic perspective, 
emphasizing that information coverage serves as a more precise and general objective than the traditional similarity and diversity.
However, directly quantifying information coverage in a combinatorial manner using information theory is challenging.
To address this, we propose a surrogate measure for information coverage in code translation, based on Abstract Syntax Tree~(AST), called the Coverage of AST~(CAST). 
%
For a given test source code and exemplar set, we extract ASTs of them while retaining only node type attributes and discarding value attributes.
The exemplar ASTs are then used to cover subtrees of the test source code AST.
CAST quantifies the proportion of nodes in the test source code AST included in the \emph{maximum covered subtrees}.
Maximizing CAST supports list-wise compositional exemplar selection, achieving ``two birds with one stone'': 
it explicitly considers syntactic structure knowledge in ICL and approximates the information coverage between the context and the test source code.
However, two new challenges arise: 
1)~How to efficiently perform subtree covering across trees, and 
2)~How to solve CAST maximization for exemplar selection, as it is an NP-hard problem that cannot be optimally solved within polynomial time.

To address these challenges, we register a unique fingerprint to each AST subtree. 
%
Fingerprints for all subtrees in an AST are efficiently extracted via a single post-order traversal, and the exemplar database can be pre-processed efficiently.
%
Next, we construct a co-occurrence matrix $M$ between the test source code and candidate exemplars, where $M_{ij}$ denotes whether the $j$-th AST subtree of the source code exists in the $i$-th candidate exemplar. 
Consequently, exemplar selection can be formulated as selecting rows from the co-occurrence matrix to maximize the $\ell_{1}$-norm of the \emph{OR} vector across selected rows.
We prove this is a standard submodular maximization problem, which has been widely studied in subset selection area.
Although this is a general NP-hard problem, we can employ a greedy algorithm that starts with an empty selection set and iteratively adds elements to maximize the marginal gain of CAST at each step until the size limit $k$ is reached.
%
This algorithm theoretically guarantees a $(1-1/e)$-approximate solution with polynomial time complexity, i.e., its performance is at least $(1-1/e)$ of the optimal solution.

To evaluate CAST, we integrate it as a plug-and-play module with state-of-the-art~(SOTA) LLMs of varying scales during inference and compare it with traditional exemplar selection strategies based on similarity and diversity. 
%
Experimental results demonstrate that our method significantly improves performance over baseline methods and outperforms other exemplar selection strategies.

Our main contributions are summarized as follows:
\begin{enumerate}
    \item 
    \textbf{Revealing meaningful insights:} 
    This paper presents two key findings: First, code syntactic knowledge, often overlooked during training, can be effectively post-incorporated into pre-trained LLMs at test time. Second, scaling model size or training data alone does not ensure the sufficient emergence of code syntactic knowledge, emphasizing the need to explicitly consider code syntactic structures into LLMs. 
    \item 
    \textbf{New objective for ICL exemplar selection, CAST:}
    We propose the first training-free and LLM-agnostic approach to incorporate code syntactic knowledge into existing LLMs during inference using ICL.
    We revisit exemplar selection in ICL from an information-theoretic perspective, showing that information coverage is a more precise and general objective.
    We introduce a surrogate measure that incorporates syntactic structure knowledge into LLMs compositionally while approximating the quantify-challenging information coverage efficiently.
    \item 
    \textbf{Practical solution for CAST maximization:}
    Although CAST maximization in exemplar selection is NP-hard, we prove it is a submodular maximization problem and propose a greedy algorithm that theoretically guarantees a $(1-1/e)$-approximate solution with practical polynomial time complexity.
    Experimental results show that CAST significantly improves the code translation performance of LLMs across various scales.
\end{enumerate}

\section{Revisiting Exemplar Selection via Information Theory for ICL in Code Translation }\label{sec:moti}


\subsection{Preliminaries of Code Translation and ICL}
Code translation can be formulated as \mbox{$\trans{}: X \mapsto Y$}, where $X$ represents the source code to be translated, $Y$ represents the target code, and $\trans{}$ denotes a translation model, such as an LLM.
For a given source code instance $X_{\text{test}}$, the translation is formally expressed as:
\begin{equation}
    \widehat{Y}_{\text{test}} = \trans(\widehat{Y}_{\text{test}}|X_{\text{test}})\,.
\end{equation}

In ICL, the database $\mathcal{D} = \{(X_i, Y_i)\}_{i=1}^n$ contains $n$ source and target code pairs. 
The indices of the database are denoted by $\mathbb{N}_n = \{1, 2, \dots, n\}$, and $S \subseteq \mathbb{N}_n$ represents a subset of these indices. 
Thus, ICL-based code translation with $k$ exemplars as context is expressed as:
\begin{equation}\label{eq:icl}
    \widehat{Y}_{\text{test}} = 
        \trans(\widehat{Y}_{\text{test}}|
            \underbrace{X_{S[1]}, Y_{S[1]}, \dots, X_{S[k]}, Y_{S[k]}}_{\text{context of $k$ paired exemplars}}, X_{\text{test}}
        )\,,
\end{equation}
where $S[i]$ represents the index of the $i$-th selected exemplar.

Therefore, the primary challenge is to select an effective context for the input source code $X_{\text{test}}$.
This requires finding a subset of indices $S$ that meets the context size constraint $|S| = k$, defined as:
\begin{equation} \label{generaleq}
    S^* = \arg\max_{S \subseteq \mathbb{N}_n, |S|=k} \mathrm{Score}(S,X_{\text{test}})\,.
\end{equation}



Traditional ICL exemplar selection relies on two main criteria: similarity and diversity~\cite{luo2024context}. 
Similarity-based methods select the top-$k$ most similar exemplars~\cite{rubin-etal-2022-learning,agrawal2022context,dalvi2022towards,wang2022towards,liu2021makes}, which are simple and fast but often yield homogeneous exemplars and sub-optimal results.
To address redundancy, some studies incorporate diversity~\cite{li2024long,li2022self,li2023mot}, but these methods typically depend on instance-wise distance metrics and neglect the compositional relationships among exemplars.
Moreover, current ICL exemplar selection methods often struggle to balance similarity and diversity,
as the trade-off is task-sensitive, limiting their robustness and generalizability.



\begin{figure*}[t]
  \centering
  \includegraphics[width=\linewidth]{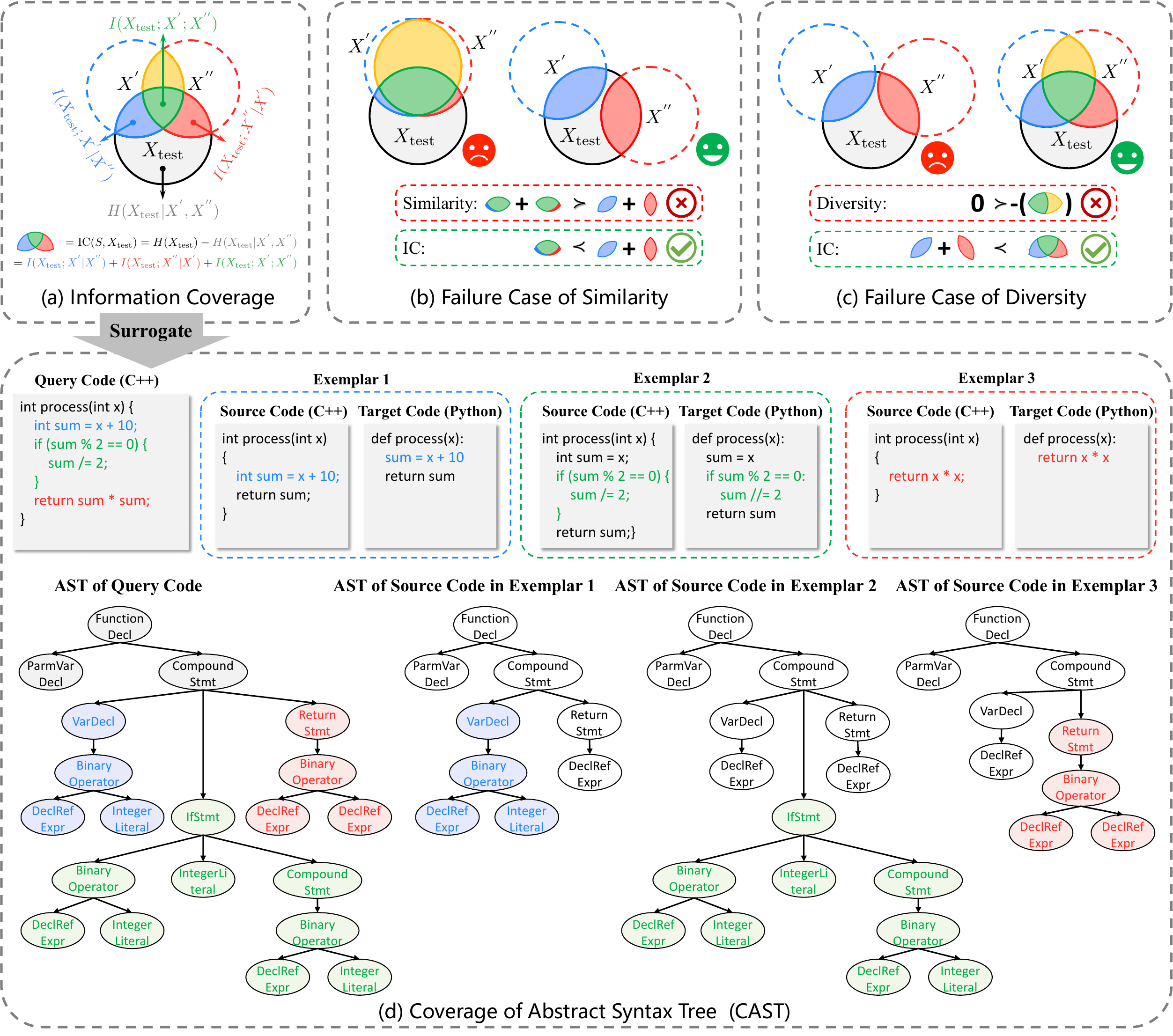}
  \caption{
      Revisiting ICL exemplar selection from an information-theoretic perspective; CAST for code translation.
    } 
  \label{fig:moti}
\end{figure*}

\subsection{Revisiting Exemplar Selection with Information Theory}

We revisit exemplar selection in ICL from an information-theoretic perspective.
In code translation, information coverage refers to 
\emph{the overlapping information between the source code of test sample and task exemplars}, 
quantifying the useful knowledge provided by exemplars for translating the current source code.
Formally, information coverage is defined as:
\begin{equation}\label{eq:ic_def}
    \mathrm{IC}(S, X_{\text{test}}) = H(X_{\text{test}}) - H(X_{\text{test}}| X_{S[1]}, \dots, X_{S[k]}) \,,
\end{equation}
where $H()$ represents information entropy.
For instance, as shown in \cref{fig:moti}~(a), if two exemplars~($k=2$) with source codes $X^{'}$ and $X^{''}$ are selected for the test source code $X_{\text{test}}$, the information coverage is:
\begin{align}\label{eq:ic_exp}
    &\mathrm{IC}(S, X_{\text{test}}) = H(X_{\text{test}}) - H(X_{\text{test}} | X^{'}, X^{''}) \\
    &\quad = I(X_{\text{test}}; X^{'} | X^{''}) + I(X_{\text{test}}; X^{''} | X^{'}) + I(X_{\text{test}}; X^{'}; X^{''})\,, \notag
\end{align}
where $I()$ denotes general mutual information, such as interaction information or conditional mutual information.
Here, the required knowledge to translate $X_{\text{test}}$~($H(X_{\text{test}})$) is divided into four parts:
$I(X_{\text{test}}; X^{'} | X^{''})$ and $I(X_{\text{test}}; X^{''} | X^{'})$ measure the individual contributions of $X^{'}$ and $X^{''}$, respectively. 
$I(X_{\text{test}}; X^{'}; X^{''})$ represents the common knowledge provided by both $X^{'}$ and $X^{''}$, 
and $H(X_{\text{test}}| X^{'}; X^{''})$ reflects the knowledge not provided by these exemplars.

Revisiting the traditional notions of similarity and diversity with the information theory, similarity can be interpreted as:
\begin{equation}
    \mathrm{Similarity}(S, X_{\text{test}}) \propto \sum_{i=1}^n I(X_{\text{test}}; X_{S[i]}) \,,
\end{equation}
which measures the total mutual information between the test sample $X_{\text{test}}$ and each selected exemplar $X_{S[i]}$.
On the other hand, pairwise diversity can be interpreted as:
\begin{equation}
    \mathrm{Diversity}(S, X_{\text{test}}) \propto -\sum_{i,j=1; i \neq j}^n I(X_{S[i]}; X_{S[j]}) \,,
\end{equation}
which accounts for the negative mutual information between selected exemplars, ensuring reduced redundancy among them.

The Venn diagram examples in~\cref{fig:moti} illustrate the failure cases of maximizing similarity or diversity individually. 
Focusing solely on similarity overemphasizes common knowledge among exemplars~(as shown in \cref{fig:moti}~(b)), resulting in redundant selections with minimal information gain. 
Conversely, optimizing diversity alone ignores the relevance between the test sample and selected exemplars and unjustly penalizes irrelevant mutual information $I(X^{'}; X^{''}|X_{\text{test}})$~(yellow part in \cref{fig:moti}~(c)).
Some methods arbitrarily combine similarity and diversity as a weighted objective. 
To compare it with information coverage~(\cref{eq:ic_exp}), consider selecting two exemplars, $X^{'}$ and $X^{''}$. 
The combined objective, weighted by $\lambda$, is given by:
\begin{align}\label{eq:combine}
    &\mathrm{Similarity}(S, X_{\text{test}}) + \lambda \cdot \mathrm{Diversity}(S, X_{\text{test}}) \notag\\
    &\qquad \propto I(X_{\text{test}}; X^{'}) + I(X_{\text{test}}; X^{''}) - \lambda I(X^{'}; X^{''}) \notag \\
    &\qquad = I(X_{\text{test}}; X^{'} | X^{''}) + I(X_{\text{test}}; X^{''} | X^{'}) \\
    &\qquad\quad + (2 - \lambda) I(X_{\text{test}}; X^{'}; X^{''}) - \lambda I(X^{'}; X^{''}|X_{\text{test}}) \,. \notag
\end{align}

It is similar to \cref{eq:ic_exp}, but there are two key differences:  
1)~It arbitrarily assigns different weights to the components of required knowledge.  
While setting $\lambda=1$ ensures fairness, increasing the selection size $k$ complicates partitioning and weighting, making the balance harder to maintain.  
2)~The term $I(X^{'}; X^{''}|X_{\text{test}})$ is unnecessary, as it represents shared knowledge of $X^{'}$ and $X^{''}$. Minimizing this term is redundant and can interfere with optimizing other valuable terms.

Consequently, information coverage serves as a more precise and general objective than similarity and diversity.
However, directly computing it, as defined in \cref{eq:ic_def}, is challenging due to the complexity of estimating the conditional distribution $\mathcal{P}(X_{\text{test}}| X_{S[1]}, \dots, X_{S[k]})$.  
To overcome this challenge, we propose a surrogate measure called CAST, which is computationally efficient and considers code syntactic knowledge.
The details are elaborated in the next section~(\cref{sec:method}).

\section{Proposed Method}\label{sec:method}

To enhance code translation in LLMs via ICL, 
we propose CAST, a surrogate measure for information coverage, 
and present a practical solution for NP-hard CAST maximization with exemplar set selection.
In \cref{sec:CAST}, we define CAST, a list-wise $\mathrm{Score}$ function based on the code's AST. 
CAST accounts for both the relevance of exemplars to the source code and the compositional relationships among multiple selected exemplars. 
In \cref{sec:solution}, we present a greedy algorithm for CAST maximization, proving that it guarantees a $(1-1/e)$-approximation for the NP-hard subset selection problem with polynomial time complexity. 
Finally, \cref{sec:overview} provides an overview of the CAST framework.

\subsection{Surrogate Scoring Function: CAST}\label{sec:CAST}

We aim to integrate code structural knowledge into LLMs via ICL. 
As discussed in \cref{sec:moti}, effective ICL exemplar selection relies on a well-designed $\mathrm{Score}$ function, with information coverage as a precise and general scoring objective. 
However, directly quantifying information coverage is challenging, and limited research has explored considering code structural knowledge in ICL.  
We propose CAST, a surrogate measure that approximates information coverage by quantifying the ratio of the \emph{maximum subtrees covered by exemplars} in the test source code AST. 
Maximizing CAST is the first training-free and model-agnostic method for explicitly post-incorporating code structural knowledge into LLMs at test time.

We employ tree-sitter\footnote{\url{https://tree-sitter.github.io/tree-sitter}} to extract the AST of the source code.
Each node contains two attributes: type and value, e.g.,
\begin{equation}\begin{aligned}
    \text{node}_1 &= \{\text{``type'': ``variable name'', ``value'': ``count''}\}\,, \\
    \text{node}_2 &= \{\text{``type'': ``function name'', ``value'': ``sort'`}\}\,.
\end{aligned}\end{equation}
The type attribute reveals syntactic information, while the value attribute captures lexical details. 
To improve generalization, we retain only the type attribute~(i.e., syntactic information) and discard the value attribute.
Let $T_{\text{test}}$ denote the AST of the current test source code~($X_{\text{test}}$), and let $T_{i}$ denote the AST of the $i$-th sample's source code~($X_{i}$) in the database. 
CAST serves as the $\mathrm{Score}$ function, a surrogate measure for information coverage with definition:
\textbf{
    A maximum covered subtree is a subtree of $T_{\text{test}}$ that exists in the AST of exemplars and contains no smaller covered subtrees. 
    CAST is computed as the ratio of nodes in the maximum covered subtrees to the total nodes in $T_{\text{test}}$.
}
The number of nodes in a tree equals the total number of its subtrees. 
Consequently, let $\mathrm{Sub}(T_S)=\mathrm{Sub}(\bigcup_{i=1}^{|S|}T_{S[i]})$ denote the subtrees of ASTs for all exemplars in the context.
CAST is then defined as:
\begin{equation}\label{eq:acr}
    \mathrm{CAST}(S, X_{\text{test}}) = \frac{ \left| 
        \mathrm{Sub}(T_{\text{test}}) \cap \mathrm{Sub}(T_S)
    \right|}{\left|
        \mathrm{Sub}(T_{\text{test}})
    \right|} \,.
\end{equation}

\begin{algorithm}[t]
    \KwIn{A tree $T$ and its root node $r$.}
    \KwOut{The fingerprint of the tree $T$.}
    $\fp \gets 0$; \\ 
    \For{$c \in \mathrm{children}(r)$}{
        $\fp \gets \mathrm{hash}(\fp + \mathrm{Fingerprint}(T, c))$;
    }
    $\fp \gets \mathrm{hash}(\fp + \mathrm{hash}(r.\mathrm{type}))$; \\    
    \Return $\fp$; \\
    \caption{$\mathrm{Fingerprint}(T, r)$}
    \label{algo:fingerprint}
\end{algorithm}

To compute $\mathrm{CAST}$ efficiently, we assign a fingerprint~($\fp$) for each subtree via hashing, as detailed in \cref{algo:fingerprint}.
The fingerprint encodes the node type~(line 5) and recursively calculates the subtree fingerprint using post-order traversal and hashing.
The set of fingerprints for the subtrees of a tree $T_i$ is denoted as $\Phi_i = \{\fp_j^{(i)}\}_{j=1}^{|T_i|}$, where all $\fp_j^{(i)}$ are computed from line 5 in \cref{algo:fingerprint} by calling $\mathrm{Fingerprint}(T, r)$ only once. 
The fingerprint set $\Phi_i$ for all samples $X_i \in \mathcal{D}$ in the database can be precomputed.
To translate a new source code $X_{\text{test}}$, 
we first compute the fingerprint set for subtrees in its AST $\Phi_{\text{test}} = \{\fp_j^{(\text{test})}\}_{j=1}^{|T_{\text{test}}|}$, 
and then construct a co-occurrence matrix $M \in \{0,1\}^{n \times |\Phi_{\text{test}}|}$, 
with each element $M_{ij}$ defined as:
\begin{equation}
    M_{ij} = \begin{cases} 
        1, & \text{if } \fp_j^{(\text{test})} \in \Phi_i \\ 
        0, & \text{otherwise} 
    \end{cases} \,.
\end{equation}
This matrix indicates whether the $j$-th subtree of the test source code’s AST~($T_{\text{test}}$) is present in the AST of the $i$-th source code sample~($T_{i}$) in the database.

To apply CAST as the $\mathrm{Score}$ function for list-wise exemplar selection in ICL~(\cref{eq:icl}), we need to maximize CAST~(\cref{eq:acr}), which is equivalent to maximizing $| \mathrm{Sub}(T_{\text{test}}) \cap \mathrm{Sub}(T_S)|$. 
Based on the definition of $M$, CAST maximization is formally expressed as:
\begin{equation}\begin{aligned}\label{eq:acr_max}
    \max_{S \subseteq \mathbb{N}_n} &\quad \left\| \bigvee_{s \in S} M[s,:] \right\|_1\\
    \text{s.t.} &\quad |S| = k \,.
\end{aligned}\end{equation}
Here, $\|\cdot \|_1$ denotes the $\ell_1$-norm of a vector, which counts the number of ones in the union vector $\bigvee_{s \in S} M[s,:] \in \{0,1\}^{|\Phi_{\text{test}}|}$.

This optimization problem is a subset selection task with a list-wise compositional objective, which is NP-hard that cannot be optimally resolve in polynomial time complexity.


\subsection{Practical Solution: Greedy Submodular Maximization}\label{sec:solution}

To address the NP-hard challenge of CAST maximization, we propose a greedy submodular maximization algorithm.
In exemplar selection for ICL, the law of diminishing marginal utility states that as more additional exemplars are selected, the marginal~(incremental) information gain is non-increasing.
This property holds for both information coverage and its surrogate measure, CAST.
Formally, this property is known as submodularity in mathematics and is defined as:
\begin{definition}[Submodularity~\cite{nemhauser1978best}]\label{thm:sub_def}
	Let $\Omega$ denote a finite set, and let $f: 2^{\Omega} \rightarrow \mathbb{R}_{\geq 0}$ be a set function, where $2^{\Omega}$ represents the power set of $\Omega$. 
    The function $f$ is defined as submodular if it satisfies:
    For every $A \subseteq B\subseteq \Omega$, and for all $x \in \Omega \setminus B$,
    \begin{equation}
        f(A \cup \{x\}) - f(A) \geq f(B \cup \{x\}) - f(B) \,,
    \end{equation}
\end{definition}
Additionally, $f$ is monotone \emph{if and only if} $f(A) \leq f(B)$ for all $A \subseteq B \subseteq \Omega$.

Maximizing a non-negative, monotone, and submodular function in subset selection referred to \emph{submodular maximization}.
Although generally NP-hard, a greedy algorithm that iteratively maximizes the marginal gain offers a theoretical $(1-1/e)$-approximate solution in polynomial time.
The surrogate CAST~(\cref{eq:acr}) corresponds to the value function \mbox{$f(S) = \left| \bigvee_{s \in S} M[s,:] \right|_1$} in the optimization problem~(\cref{eq:acr_max}).
This value function is non-negative, monotone, and submodular. 
We present a lemma to prove these properties:
\begin{lemma}[Properties of $\ell_1$-norm in boolean vector operations]\label{lemma:mag_l1_norm}
    Let \( A, B \in \{0,1\}^m \). Then, the following holds:
    \begin{equation}
        \| B \wedge \neg A \|_1 = \| A \vee B \|_1 - \| A \|_1 = \| B \|_1 - \| B \wedge A \|_1 \,,
    \end{equation}
    where $\| \cdot \|_1$ denotes the $\ell_1$-norm.

    \begin{proof}
        The $\ell_1$-norm of a binary vector \(A \in \{0,1\}^m\) counts the number of elements equal to ``$1$'':
        \begin{equation}
            \| A \|_1 = \sum_{i=1}^m A_i \,.
        \end{equation}
        Boolean operations on binary vectors are performed element-wise, as follows:
        \begin{align}
            (B \wedge \neg A)_i &= B_i (1 - A_i) = B_i - (B \wedge A)_i \,, \\
            (A \vee B)_i &= A_i + B_i - (A_i \wedge B_i) = A_i + (B \wedge \neg A)_i \,. \notag
        \end{align}
        Therefore,
        \begin{align}
            \| B \wedge \neg A \|_1 &= \sum_{i=1}^m B_i - (B \wedge A)_i = \| B \|_1 - \| B \wedge A \|_1 \,, \notag \\
            \| B \wedge \neg A \|_1 &= \sum_{i=1}^m (A \vee B)_i - A_i = \| A \vee B \|_1 - \| A \|_1 
        \end{align}
        Consequently, the proof is complete.
    \end{proof}
\end{lemma}

\begin{proposition}[The value function for $\mathrm{CAST}$ maximization is non-negative, monotone, and submodular.]
    Given a binary matrix \( M \in \{0,1\}^{n \times m} \), the value function for an index set \( S \subseteq \mathbb{N}_n \),
    \begin{equation}
        f(S) = \left\| \bigvee_{s \in S} M[s,:] \right\|_1\,,
    \end{equation}
    is non-negative, monotone, and submodular.
    
    \begin{proof}
    The function \( f(S) = \left\| \bigvee_{s \in S} M[s,:] \right\|_1 \) computes the number of ones resulting from the bitwise \emph{OR} operation applied across the rows indexed by \( S \).

    1) \textbf{Non-negativity}:
        Since \( f(S) \) counts the number of ones in the bitwise \emph{OR} operation, it is always non-negative. Thus, we have \( f(S) \geq 0 \) for all \( S \subseteq \mathbb{N}_n \).

    2) \textbf{Monotonicity}:
        For any subsets \( A \subseteq B \subseteq \mathbb{N}_n \), adding more rows to the bitwise \emph{OR} operation cannot decrease the count of ones. Hence, \( f(A) \leq f(B) \), i.e., \( f(S) \) is monotone.

    3) \textbf{Submodularity}:
        Given a subset \( S \subseteq \mathbb{N}_n \), when adding a new index \( x \in \mathbb{N}_n \setminus S \), the marginal gain is given by:
        \begin{equation}\begin{aligned}
            \Delta_{S}(x) &= f(S\cup \{x\}) - f(S) \\
                &= \left\| \bigvee_{s \in S\cup\{x\}} M[s,:] \right\|_1 -  \left\| \bigvee_{x \in S} M[s,:] \right\|_1 \\
                &= \left\| M[x,:] \wedge \neg \bigvee_{s \in S} M[s,:] \right\|_1 \,.
        \end{aligned}\end{equation}
        This follows from $\| A \vee B \|_1 - \| A \|_1 = \| B \wedge \neg A \|_1$ in Lemma~\ref{lemma:mag_l1_norm}.
        Hence, for any subsets \( A \subseteq B \subseteq \mathbb{N}_n \):
        \begin{align}
            \Delta_{B}(x) 
            &= \left\| M[x,:] \wedge \neg \bigvee_{s \in B} M[s,:] \right\|_1 \notag \\
            &= \left\| M[x,:] \wedge \neg \bigvee_{s \in A} M[s,:] \wedge \neg \bigvee_{s \in B\setminus A} M[s,:] \right\|_1 \notag \\
            &= \underbrace{
                \left\| M[x,:] \wedge \neg \bigvee_{s \in A} M[s,:] \right\|_1
            }_{\Delta_{A}(x)} \\
            &\quad - \underbrace{
                \left\| M[x,:] \wedge \neg \bigvee_{s \in A} M[s,:] \wedge \bigvee_{s \in B\setminus A} M[s,:] \right\|_1 
            }_{\geq 0} \,, \notag
        \end{align}
        where the last equation follows from Lemma~\ref{lemma:mag_l1_norm}, which states that $\| B \wedge \neg A \|_1 = \| B \|_1 - \| B \wedge A \|_1$.
        So, \( \Delta_{A}(x) -  \Delta_{B}(x) \geq 0 \), thereby the value function \( f \) is submodular,
        $\forall  A \subseteq B \subseteq \mathbb{N}_n$ and $\forall x \in \mathbb{N}_n \setminus B$, the following holds:
        \begin{equation}
             f(A \cup \{x\}) - f(A) \geq f(B \cup \{x\}) - f(B).
        \end{equation} 
        Hence, the function \( f(S) \) is non-negative, monotone, and submodular. 
        The proof is complete.
        \end{proof}
\end{proposition}
Consequently, the list-wise exemplar set selection problem in \cref{eq:acr_max} is a \emph{submodular maximization} problem.

Submodular maximization is generally NP-hard. However, when the value function is non-negative, monotone, and submodular, a greedy algorithm can be used. It starts with an empty set and iteratively adds the element with the maximum marginal gain~($\Delta_S(x)$): 
\begin{equation}\label{eq:gen_greedy_sub_max}
    x^* = \arg\max_{x \in \Omega \setminus S} f(S \cup \{x\}) - f(S)\,,
\end{equation}
continuing until \( k \) elements are selected.
It provides a solution within polynomial time and guarantees a near-optimal approximation, as detailed in the following theorem~\cite{nemhauser1978best,sunh_mdp3}:
\begin{theorem}[ $(1-1/e)$-approximation of greedy submodular maximization~\citet{nemhauser1978best}]\label{theo:approx}
    For any non-negative, monotone submodular function \( f \), the greedy heuristic produces a solution \( \hat{S} \) of size \( |\hat{S}| = k \) with the following approximation guarantee:
    \begin{equation}\begin{aligned}
        f(\hat{S}) 
            &\geq \left( 1 - \left(1 - \frac{1}{k}\right)^k \right) f(S^*) \\
            &\geq \lim_{k \to +\infty} \left( 1 - \left(1 - \frac{1}{k}\right)^k \right) f(S^*) \\
            & = \left(1 - \frac{1}{e}\right) f(S^*)\,,
    \end{aligned}\end{equation}
    where $S^*$ represents the optimal solution to the submodular maximization, defined as \mbox{$S^* = \arg\max_{S 
    \subseteq \Omega, |S| \leq k} f(S)$.}
\end{theorem}

Therefore, we propose a greedy algorithm to solve the CAST maximization problem, guided by \cref{eq:gen_greedy_sub_max}.
The algorithm starts with an empty selected set, \( S \leftarrow \emptyset \). 
In each iteration, the algorithm selects a new exemplar that maximizes the marginal gain \( \Delta_S(f) \) of 
\cref{eq:acr_max}:
\begin{equation}\begin{aligned}\label{eq:marginal_max}
    j &= \arg\max_{i \in \mathbb{N}_n \setminus S} 
        \left\| \bigvee_{s \in S\cup\{i\}} M[s,:] \right\|_1 -  \left\| \bigvee_{s \in S} M[s,:] \right\|_1 \,, \\
      &\qquad\qquad \Downarrow \text{\textbf{Lemma~\ref{lemma:mag_l1_norm}}} \\
    j &= \arg\max_{i \in \mathbb{N}_n \setminus S} 
        \left\| M[i,:] \wedge \neg \bigvee_{s \in S} M[s,:] \right\|_1 \,.
\end{aligned}\end{equation}
The selected index \( j \) is added to \( S \) in each iteration until the subset reaches the predefined size limit \( k \). 
The pseudocode of this greedy CAST maximization is provided in \cref{algo:greedy}.


\begin{algorithm}[t]
    \KwIn{
        Co-occurrence matrix \( M \in \{0,1\}^{n \times |\Phi_{\text{test}}|} \) and selection capacity limit \( k \ll n \).
    }
    \KwOut{
        Subset of exemplar indices \( S \subseteq \mathbb{N}_n \) such that \( |S| = k \).
    }
    Initialize the selected set: \( S \leftarrow \emptyset \); \\
    Initialize the coverage mask: \( c \leftarrow \mathbf{0}^{|\Phi_{\text{test}}|} \); \\
    \For{\( i \leftarrow 1 \) \textbf{to} \( k \)}{
        \( s \leftarrow -1 \); \( \text{max\_gain} \leftarrow -1 \); \\
        \For{\( j \leftarrow 1 \) \textbf{to} \( n \)}{
            \If{\( j \notin S \)}{
                $g \leftarrow \| M[j,:] \wedge \neg c \|_1$; \tcp{\small{marginal gain}}
                \If{\( g > \text{max\_gain} \)}{
                    \( \text{max\_gain} \leftarrow g \); \\
                    \( s \leftarrow j \);
                }
            }
        }
        Update coverage mask: \( c \leftarrow c \vee M[s,:] \); \\
        Update selected set: \( S \leftarrow S \cup \{s\} \); \\
    }
    \Return \( S \);
    \caption{\mbox{Greedy CAST Submodular Maximization}}
    \label{algo:greedy}
\end{algorithm}

As shown in Theorem~\ref{theo:approx}, our CAST maximization method in \cref{eq:marginal_max} and \cref{algo:greedy} provides a \((1 - 1/e)\)-approximation, ensuring sufficient effectiveness. 
For instance, if \( k = 3 \) exemplars are selected, the approximation ratio reaches at least \mbox{\( 1 - (1 - 1/3)^3 \approx 70.37\% \).}

The time complexity of \cref{algo:greedy} includes three parts: 
$\mathcal{O}(k)$ for the loop in line 3, 
$\mathcal{O}(n)$ for the loop in line 5, 
and $\mathcal{O}(|\Phi_{\text{test}}|)$ for bitwise boolean computations in line 7. 
Thus, the total time complexity of the CAST algorithm is $\mathcal{O}(nk \cdot |\Phi_{\text{test}}|)$. 
Let \( L_{X_{\text{test}}} \) denote the length of the test source code. 
The size of the subtree fingerprint set is upper-bounded by \( |\Phi_{\text{test}}| \leq L_{X_{\text{test}}} \log(L_{X_{\text{test}}}) \).
Moreover, the selection size \( k \) and the subtree scale \( |\Phi_{\text{test}}| \) are much smaller than the database size \( n \), i.e., \( k \ll n \) and \( |\Phi_{\text{test}}| \ll n \).
As a result, the greedy CAST maximization algorithm in \cref{algo:greedy} is computationally efficient.
Furthermore, the iterations in lines 5-13 are independent, allowing the loop in line 5 to be parallelized for faster computation.
With sufficient resources in industrial applications, this optimization reduces the time complexity to $\mathcal{O}(k \cdot |\Phi_{\text{test}}|)$, further improving efficiency.

Constructing the co-occurrence matrix takes \( \mathcal{O}(n \cdot |\Phi_{\text{test}}|) \) time, based on its size~($M^{n \times |\Phi_{\text{test}}|)}$). 
Extracting the subtree fingerprint set \( \Phi_{\text{test}} \) from the test source code takes \( \mathcal{O}(|\Phi_{\text{test}}|) \) time using a single post-order traversal on \( T_{\text{test}} \).
Moreover, the subtree fingerprint sets for database samples can be precomputed, avoiding extra computations at test time.
An inverted index links each subtree fingerprint \( \mathrm{fp} \) to the corresponding database samples. 
For a new test source code, the inverted index~(e.g., using ElasticSearch) efficiently retrieves co-occurrence coordinates for each \( \mathrm{fp} \in \Phi_{\text{test}} \) in \( \mathcal{O}(|\Phi_{\text{test}}|) \) time.
Each \( \mathrm{fp} \) retrieves a sparse column~(in \( \{0,1\}^n \)) for the co-occurrence matrix \( M \in \{0,1\}^{n \times |\Phi_{\text{test}}|} \). 
Combining sparse columns to form the full matrix \( M \) takes \( \mathcal{O}(n \cdot |\Phi_{\text{test}}|) \).


In summary, constructing the co-occurrence matrix for CAST takes \(\mathcal{O}(n \cdot |\Phi_{\text{test}}|)\), while the greedy CAST algorithm in \cref{algo:greedy} has a polynomial time complexity of \(\mathcal{O}(nk \cdot |\Phi_{\text{test}}|)\). 
Using distributed and parallel computation, this can be further reduced to \(\mathcal{O}(k \cdot |\Phi_{\text{test}}|)\). 
Thus, our approach ensures both theoretical effectiveness with a near-optimal \((1 - 1/e)\)-approximation and practical efficiency.

\begin{figure*}[!t]
    \centering
    \includegraphics[width=\linewidth]{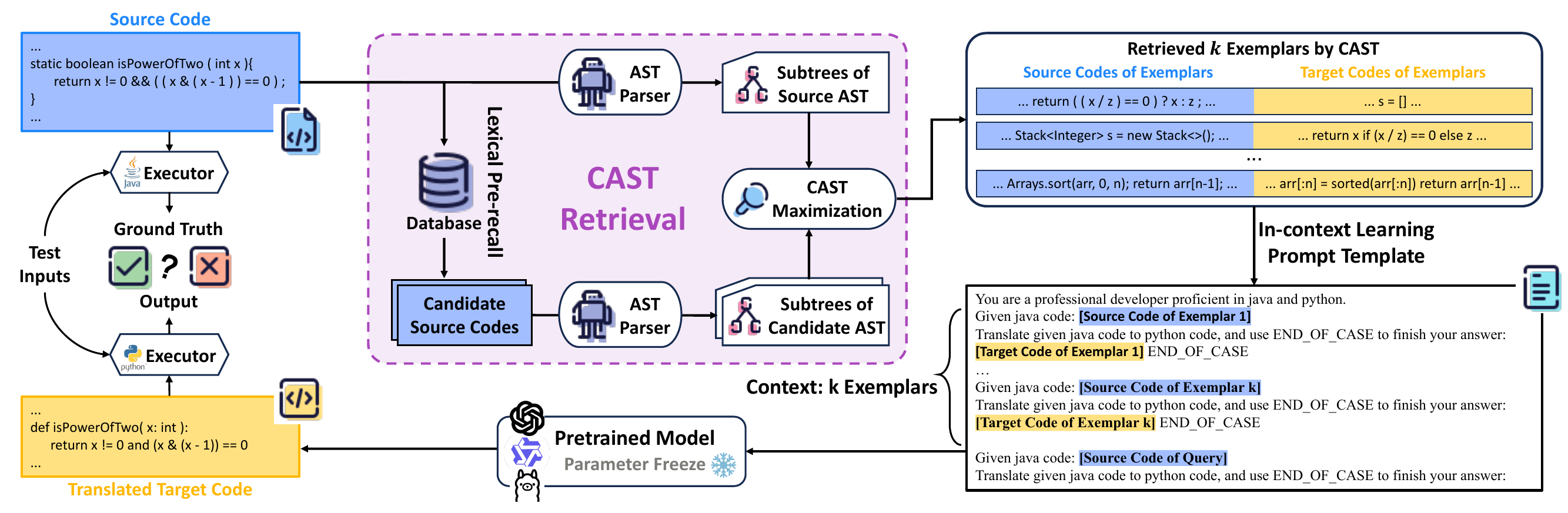}
    \caption{
        Overview of post-incorporating code syntactic knowledge into pre-trained models using ICL based on CAST retrieval.
    }
    \label{fig:framework}
    \vspace{-0.5cm}
\end{figure*}

\subsection{Overall Framework of CAST}\label{sec:overview}


\Cref{fig:framework} illustrates the overall pipeline of our approach, which post-incorporates code syntactic structure knowledge into off-the-shelf LLMs through ICL based on Covering Abstract Syntax Tree~(CAST) retrieval.
To accelerate the overall process, lexical similarity is used to coarsely pre-recall $\lfloor t \cdot k \rfloor$ candidate samples, where \( t \geq 1 \) is a hyperparameter.
Specifically, the Levenshtein Distance is employed to recall the top-$\lfloor t \cdot k \rfloor$ nearest candidates.
The fingerprint set of the test source code $\Phi_{\text{test}}$ is extracted using \cref{algo:fingerprint}, and a co-occurrence matrix $M$ is constructed with all pre-recalled candidate samples.
Then, \cref{algo:greedy} selects \( k \) exemplars as context.
The selected $k$ exemplar translation pairs are directly concatenated with the test source code to create the prompt, as shown in \cref{fig:framework}.
This prompt is then input into pre-trained models, such as LLMs, to translate the test source code.

Besides, our CAST retrieval method includes two variants: 
1)~\textbf{CAST-F}: Selects a fixed number of exemplars based on the predefined capacity limit \( k \). 
2)~\textbf{CAST-A}: Automatically determines the number of exemplars using a CAST~(defined in \cref{eq:acr}) threshold, allowing early stop when the coverage rate reaches the threshold.


\section{Experiments}

To evaluate CAST, we integrate it as a plug-and-play module with SOTA LLMs of varying scales during inference and compare it with traditional exemplar selection strategies based on similarity and diversity. 
Extensive experiments are conducted to evaluate the effectiveness and efficiency of CAST on thousands of translation tasks.

\subsection{Dataset and Database}
We use the dataset from~\citet{transcoder1}, containing 948 parallel functions in Python, Java, and C++, widely used in code translation~\cite{transcoder1,transcoder2,transcoder3}. 
Of these, 568 include unit tests for at least one programming language: 464 for Python, 482 for Java, and 467 for C++. 
Following the previous work~\cite{yang2024exploring}, our experiments focus on this subset. 
\textcolor{blue}{
To further validate our approach, we evaluate various methods on AVATAR~\cite{ahmad2023avatar}, a more realistic benchmark constructed from competitive programming sites, online platforms, and open-source repositories.
We also extend our evaluation beyond code translation to code summarization using benchmarks from CodeXGLUE~\cite{CodeXGLUE}.
}
%
%
ICL requires a separate database. 
For each pre-trained LLM, we use it to translate the entire test dataset and verify the predictions using test input cases. 
Samples that pass all test input cases are added to the database for this LLM. 
Models are evaluated on the remaining data, and final metrics are computed using the entire dataset, treating the database samples as exactly correct.
\textcolor{dgreen}{The replication package is publicly available at \url{https://doi.org/10.5281/zenodo.16920815}.}

\subsection{Baselines}
We apply CAST-F and CAST-A to SOTA LLMs and compare them with LLMs using various ICL exemplar selection strategies, as well as several learning-based translation models.

\begin{table*}[t]
    \centering
    \setlength\tabcolsep{1em}
    \caption{
        Results of pre-trained models, including LLMs and code translation models, with one-shot fixed exemplar strategy.
    }
    \label{tab:baselines}
    \begin{threeparttable}
    \begin{tabular}{lcccccccccc}
        \toprule
            \multirow{2.5}{*}{\textbf{Models}} & 
            \multicolumn{2}{c}{\textbf{C++ to Python}} & \multicolumn{2}{c}{\textbf{Python to C++}} & \multicolumn{2}{c}{\textbf{Java to C++}} & \multicolumn{2}{c}{\textbf{C++ to Java}} & \multicolumn{2}{c}{\textbf{Average}} \\           
            
            \cmidrule(lr){2-3} \cmidrule(lr){4-5} \cmidrule(lr){6-7} \cmidrule(lr){8-9} \cmidrule{10-11}
            
            \multicolumn{1}{c}{} & 
            \textbf{CA} & \textbf{EM} & 
            \textbf{CA} & \textbf{EM} &
            \textbf{CA} & \textbf{EM} &
            \textbf{CA} & \textbf{EM} &
            \textbf{CA} & \textbf{EM} \\
        \midrule
            \multicolumn{11}{c}{\textbf{Off-the-shelf Large Language Models}} \\ 
      \midrule
            CodeGen-6B & 
            30.17\% & 4.40\% & 22.70\% & 1.94\% & 35.12\% & 7.57\% & 11.20\% & 2.46\% & 24.80\% & 4.09\% \\
            CodeLLaMA-13B &
                23.24\% & 2.29\% & 16.90\% & 2.11\% & 11.62\% & 3.35\% & 16.73\% & 3.52\% & 14.48\% & 2.82\% \\
            CodeLLaMA-34B & 
                45.07\% & 6.16\% & 39.08\% & 4.23\% & 23.59\% & 7.57\% & 38.38\% & 14.96\% & 36.53\% & 8.23\% \\
            CodeLLaMA-70B & 44.01\% & 4.58\% & 50.00\% & 3.70\% & 49.65\% & 16.37\% & 43.66\% & 11.80\% & 46.83\% & 9.11\% \\
            DeepSeek-Coder-6.7B &
                44.97\% & 7.22\% & 19.75\% & 3.14\% & 34.04\% & 13.73\% & 53.17\% & 17.78\% & 37.98\% & 10.47\% \\
            DeepSeek-Coder-33B & 
                53.70\% & 7.75\% & 28.87\% & 4.01\% & 39.44\% & 14.96\% & 54.06\% & 18.31\% & 44.02\% & 11.26\% \\
      
            Gemma2-2B & 47.36\% & 6.51\% & 39.61\% & 2.64\% & 54.05\% & 12.15\% & 22.18\% & 5.28\% & 40.80\% & 6.65\% \\
            Gemma2-9B &
                 60.39\% & 8.45\% & 59.26\% & 4.75\% & 65.61\%  & \textbf{31.34\%}  & 20.81\%  &  23.06\% &  51.52\% & \textbf{16.90\%} \\ 
            Gemma2-27B & 
                63.38\% & 8.45\% & 63.91\% & 5.33\% & 68.49\% & 24.96\% & \textbf{68.31\%} & 21.83\% & 66.02\%	& 15.14\% \\ 

            Qwen2-7B &
                58.27\% & 6.16\% & 50.70\% & 4.40\% & 60.74\% & 13.20\% & 50.88\% & 11.62\% & 55.15\% & 8.85\% \\ 
            Qwen2.5-72B &
                69.01\% & 8.63\% & 68.49\% & 6.34\% & 71.30\% & 23.42\% & 64.61\% & 24.30\% & 68.36\% & 15.67\% \\
      

            LLaMA3-70B &
                69.37\% & 5.63\% & 60.74\% & \textbf{10.03\%} & 67.96\% & 23.39\% & 64.61\% & \textbf{27.99\%} & 65.67\% & 16.76\% \\ 
        
            GPT-4o-mini  & 70.95\%  & 9.33\%  & \textbf{69.01\%}  & 6.16\%  &  70.95\% & 6.69\%  & 54.58\%  & 22.01\%  &  66.37\% & 11.05\%\\
            GPT-4o  & \textbf{71.65\%}  &  \textbf{9.49\%} &  68.66\% &  5.99\% & \textbf{72.71\%}  & 23.02\%  & 65.85\%  &  26.41\% &  \textbf{69.72\%} & 16.23\% \\
            
        \midrule
            \multicolumn{11}{c}{\textbf{Learning-based Code Translation Models}}\\ 
        \midrule
            TransCoder &
                36.64\% & 2.29\% & 30.40\% & 0.88\% & 27.84\% & 5.81\% & 49.77\% & 14.39\% & 36.16\% & 5.84\% \\
            TransCoder-IR & 
                / & / & / & / & 40.99\% & 14.26\% & 50.53\% & 18.28\% & 45.76\% & 16.27\% \\
            TransCoder-ST &
                46.34\% & 2.64\% & 47.85\% & 1.06\% & 49.68\% & 9.68\% & 64.73\% & 17.43\% & 52.15\% & 7.70\%\\ 
        \bottomrule        
    \end{tabular}
    \begin{tablenotes}
        \item[$\Phi$] \scriptsize For clarification, we adopt ``-'' to concatenate LLM and corresponding parameter size for discrimination. For example, Qwen2 with 7B parameters is dubbed Qwen2-7B.
    \end{tablenotes}
    \end{threeparttable}
\end{table*}

\subsubsection{Baseline Code Translation Models}
We compare our method with recent LLM families across various parameter scales, consisting of 14 advanced LLMs:
\begin{itemize}
    \item \textbf{CodeGen}~\cite{nijkamp2022codegen}: 
        An LLM trained in natural language and programming data for conversation-based program synthesis. 
        We employ CodeGen-6B-Multi in the experiment.
    \item \textbf{Qwen}~\cite{qwen2,yang2024qwen2}: 
        A suite of models with parameter sizes ranging from 0.5B to 72B, including dense and Mixture-of-Experts variants, excelling in language, multilingual tasks, coding, math, and reasoning.
    \item \textbf{Gemma}~\cite{team2024gemma}: 
        A lightweight LLMs, ranging in scale from 2B to 27B parameters. 
    \item \textbf{DeepSeek-Coder}~\cite{guo2024deepseek,zhu2024deepseek}:
        A series of open-source code LLMs, pre-trained on 2 trillion tokens using the fill-in-the-blank task to enhance code generation and infilling. 
     \item \textbf{LLaMA}~\cite{touvron2023llama,Hugo:2023llama2}: 
        Multilingual LLMs with parameter sizes from 8B to 70B, are designed to enhance inference, code generation, and instruction compliance. 
     \item \textbf{CodeLLaMA}~\cite{codellama}: 
        A series of code-centric LLMs derived from LLaMA2~\cite{Hugo:2023llama2}, further trained on 500B code tokens.
    \item \textcolor{blue}{\textbf{GPT-4o}~\cite{gpt4}:
     Advanced generative AI models developed by OpenAI. Although not explicitly trained for code generation, they demonstrate notable performance in this domain. Their effectiveness in handling code generation tasks is primarily attributed to their massive scale. 
}
\end{itemize}
%
Moreover, three state-of-the-art learning-based code translation models serve as baselines. 
\begin{itemize}
    \item \textbf{TransCoder}~\cite{transcoder1}: 
        An unsupervised code translation model using monolingual samples, pre-trained with cross-lingual modeling, denoising, and back-translation.
    \item \textbf{TransCoder-IR}~\cite{transcoder2}: 
        A model extended on TransCoder that incorporates compiler intermediate representation~(IR) to enhance code translation. 
    \item \textbf{TransCoder-ST}~\cite{transcoder3}: 
        A model extended on TransCoder that utilizes automatically generated test cases to filter out invalid translation results.
\end{itemize}
 
\subsubsection{Baseline ICL Exemplar Selection Strategies}
We evaluate several baseline ICL exemplar selection strategies to compare with our proposed CAST variants. 

\textcolor{blue}{These strategies encompass simple heuristic methods, similarity-based approaches (e.g., LD, CodeBERT, and BM25), and diversity-oriented techniques (e.g., Diversity), which are essentially different implicit surrogates of \cref{eq:combine}. Considering only similarity is equivalent to setting $\lambda=0$ in \cref{eq:combine}.}
\textcolor{blue}{Each exemplar selection method can be defined as a specific instantiation of the general formulation~\cref{generaleq}.}

{\color{blue}
  \textbf{Fixed Exemplars (Fixed):} A static exemplar set is shared across all test samples:
  \begin{equation}
  S^* = S_{\text{fixed}}, \quad \forall X_{\text{test}}\,.
  \end{equation}
  
\textbf{Random Sampling (Random):} $k$ exemplars are uniformly sampled from the candidate pool, independent of $X_{\text{test}}$:
  \begin{equation}
  S^* \sim \text{Uniform}(\mathbb{N}_n), \quad |S^*| = k\,.
  \end{equation}

\textbf{Retrieval by Levenshtein Distance (LD):} The score is defined as the negative sum of Levenshtein distances between each exemplar and the test sample:
  \begin{equation}
  S^* = \arg\max_{S \subseteq \mathbb{N}_n,\, |S| = k} -\sum_{i \in S} \text{LD}(X_i, X_{\text{test}})\,.
  \end{equation}
\textcolor{dgreen}{Similarly, the method \textbf{Retrieval by AST Edit Distance (AST-ED)} is defined based on the Tree Edit Distance~\cite{zhang1989ted} between each exemplar and the test sample, where only the type information of AST nodes is retained for distance calculation to improve generalization, aligning with the setting of CAST.}

\textbf{Retrieval by CodeBERT Similarity (CodeBERT):} Each sample is encoded using a CodeBERT embedding function $f_{\text{CB}}(\cdot)$, and cosine similarity is used:
  \begin{equation}
  S^* = \arg\max_{S \subseteq \mathbb{N}_n,\, |S| = k} \sum_{i \in S} \cos\left(f_{\text{CB}}(X_i), f_{\text{CB}}(X_{\text{test}})\right)\,.
  \end{equation}

\textbf{Retrieval by BM25 (BM25):} The score is the sum of BM25 relevance scores between each candidate and the test input:
  \begin{equation}
  S^* = \arg\max_{S \subseteq \mathbb{N}_n,\, |S| = k} \sum_{i \in S} \text{BM25}(X_i, X_{\text{test}})\,.
  \end{equation}

\textbf{Diversity with Clustering (Diversity):} The candidate pool is clustered into $k$ groups using CodeBERT embeddings, and the most similar sample in each cluster is selected:
  \begin{equation}
  \mathbb{N}_n \xrightarrow[]{\text{k-means}(f_{\text{CB}}(X_i))} \{C_1, \dots, C_k\}\,,
  \end{equation}
  where $C_j$ represents the index set of the $j$-th cluster.
  \begin{equation}
    S^*_{[i]} = \arg\max_{j \in C_i} \cos\left(f_{\text{CB}}(X_j), f_{\text{CB}}(X_{\text{test}})\right)\,,
  \end{equation}
  for $i \in \{1,\dots, k\}$, and $S^* = S^*_{[1:k]}$.
}

\subsection{Evaluation Metrics}
We employ Computational Accuracy~(CA) and Exact Match Accuracy~(EM) to evaluate model performance, following the prior works~\cite{transcoder1,transcoder2,transcoder3,yang2024exploring}.

Traditional studies primarily use the Bi-Lingual Evaluation Understudy (BLEU)~\cite{papineni2002bleu} metric for evaluating code translation as a natural language translation task.
However, in code translation, correctness of code execution is more critical.
CA measures the percentage of translated code that produces identical execution results to the ground truth code under the same inputs.
This metric assesses execution consistency, assuming the provided inputs adequately cover all boundary conditions, and is defined as:
\begin{equation}
    \mathrm{CA}\triangleq 
    \mathop{\mathbb{E}}\limits_{X_\text{test}}\left[
        \mathbb{I} \left(
                \mathrm{Exec}(X_{\text{test}}, \mathcal{D}_X^{\text{test}})=\mathrm{Exec}(\widehat{Y}_\text{test}, \mathcal{D}_X^{\text{test}})
            \right)
        \right]
\end{equation}
$\mathcal{D}_X^{\text{test}}$ denotes the input set for the current source code $X_\text{test}$.
$\mathrm{Exec}(X_\text{test}, \mathcal{D}_X^{\text{test}})$ and 
$\mathrm{Exec}(\widehat{Y}_\text{test}, \mathcal{D}_X^{\text{test}})$ indicate the outputs produced by executing the source code $X_\text{test}$ and the translated target code $\widehat{Y}_\text{test}$ with the input cases $\mathcal{D}_X^{\text{test}}$, respectively.

Additionally, since the datasets provide ground truth code in the target programming languages, we evaluate EM accuracy, defined as the proportion of target translations that exactly match the ground truth target code.

\subsection{Empirical Results and Analysis}

\begin{table*}[t]
    \setlength\tabcolsep{1.12em}
    \centering
    \caption{
        Results of LLMs of various parameter scales for 5-shot Code Translation upon different ICL strategies.
    }
    \label{tab:main_res}
    \begin{tabular}{llccccccl}
    \toprule
        \textbf{Models} & \textbf{Strategy} & \textbf{C++ to Py} & \textbf{Py to C++} & \textbf{Java to C++} & \textbf{C++ to Java} & \textbf{Py to Java} & \textbf{Java to Py} & \textbf{Average} \\
    \midrule
        \multirow{7}{*}{Qwen2-7B} 
            & Random & 
                63.38\% & 59.31\% & 66.54\% & 61.27\% & 49.74\% & 67.43\% & ~61.28\%$^\bullet$\\
           
            & LD & 
                64.96\% & 57.69\% & 67.96\% & 63.73\% & 52.03\% & 67.61\% & ~62.33\%$^\bullet$\\
            & \color{dgreen}AST-ED & \color{dgreen}64.96\% & \color{dgreen}61.09\% & \color{dgreen}68.13\% & \color{dgreen}64.44\% & \color{dgreen}52.99\% & \color{dgreen}67.08\% & ~\color{dgreen}63.12\%$^\bullet$ \\
            & \color{blue}BM25 & \color{blue}64.61\% & \color{blue}58.98\% & \color{blue}67.25\% & \color{blue}63.73\% & \color{blue}50.18\% & \color{blue}67.78\% & ~\color{blue}62.09\%$^\bullet$\\
            & CodeBERT & 
                64.79\% & 57.02\% & 66.02\% & 63.91\% & 53.35\% & 66.37\% & ~61.91\%$^\bullet$\\
            & Diversity & 64.96\%
                 & 59.51\% & 67.42\% & 65.15\% & 51.41\% & 67.61\% & ~62.68\%$^\circ$ \\
            & \textbf{CAST-F} & 
                \textbf{65.32\%} & \textbf{61.62\%} & \textbf{68.84\%} & \textbf{65.15\%} & \textbf{55.77\%} & \textbf{69.01\%} & ~\textbf{64.28\%}~ \\ 


    \midrule
        \multirow{7}{*}{Gemma2-27B}
            & Random &
                67.25\% & 67.08\% & 70.77\% & 71.83\% & 66.02\% & 69.89\% & ~68.81\%$^\bullet$\\
            & LD & 
                67.43\% & 66.73\% & 70.95\% & 69.54\% & 65.49\% & 69.54\% & ~68.28\%$^\bullet$\\
            & \color{dgreen}AST-ED & \color{dgreen}67.25\% & \color{dgreen}67.96\% & \color{dgreen}70.77\% & \color{dgreen}69.72\% & \color{dgreen}66.55\% & \color{dgreen}69.19\% & ~\color{dgreen}68.57\%$^\bullet$ \\
            & \color{blue}BM25 & \color{blue}68.13\% & \color{blue}67.43\% & \color{blue}70.95\% & \color{blue}71.65\% & \color{blue}65.32\% & \color{blue}69.54\% & ~\color{blue}68.84\%$^\bullet$\\
            & CodeBERT & 
                67.25\% & 66.55\% & 70.42\% & 70.25\% & 64.79\% & 69.72\% & ~68.16\%$^\bullet$\\
            & Diversity & 68.13\% & 66.55\% & 70.77\% & 69.37\% & 64.08\% & 69.89\% & ~68.13\%$^\bullet$\\
            & \textbf{CAST-F} & 
                \textbf{70.60\%} & \textbf{68.66\%} & \textbf{71.13\%} & \textbf{72.49\%} & \textbf{66.90\%} & \textbf{70.40\%} & ~\textbf{70.03\%}~ \\ 

    \midrule
        \multirow{7}{*}{Qwen2.5-72B} 
            & Random &
                69.54\% & 70.07\% & 73.94\% & 66.73\% & 61.80\% & 72.01\% & ~69.02\%$^\circ$  \\
            & LD &
                70.25\% & 70.42\% & 71.48\% & 70.95\% & 59.15\% & 72.18\% & ~69.07\%$^\circ$  \\
            & \color{dgreen}AST-ED & \color{dgreen}70.60\% & \color{dgreen}70.07\% & \color{dgreen}73.24\% & \color{dgreen}66.90\% & \color{dgreen}63.03\% & \color{dgreen}71.65\% & ~\color{dgreen}69.25\%$^\circ$ \\
            & \color{blue}BM25 & \color{blue}70.42\% & \color{blue}70.07\% & \color{blue}70.95\% & \color{blue}68.31\% & \color{blue}64.96\% & \color{blue}71.13\% & ~\color{blue}69.31\%$^\circ$  \\
            & CodeBERT & 
                70.42\% & 69.72\% & 71.65\% & 70.42\% & 60.56\% & 72.01\% & ~69.13\%$^\bullet$\\
            & Diversity & 70.25\% & 69.89\% & 73.06\% & 71.30\% & 64.96\% & 71.65\% & ~70.19\%$^\bullet$\\
            & \textbf{CAST-F} & 
                \textbf{70.77\%} & \textbf{70.77\%} & \textbf{75.53\%} & \textbf{73.06\%} & \textbf{65.31\%} & \textbf{72.54\%} & ~\textbf{71.33\%}~ \\ 
    \midrule
        \multirow{7}{*}{\color{blue}GPT-4o} 
            & \color{blue}Random & \color{blue}72.01\% & \color{blue}69.19\% & \color{blue}73.42\% & \color{blue}70.95\% & \color{blue}69.01\% & \color{blue}71.48\% & ~\color{blue}71.01\%$^\bullet$\\
            & \color{blue}LD & \color{blue}72.36\% & \color{blue}70.25\% & \color{blue}73.59\% & \color{blue}71.48\% & \color{blue}69.54\% & \color{blue}71.48\% & ~\color{blue}71.45\%$^\bullet$ \\
            & \color{dgreen}AST-ED & \color{dgreen}72.71\% & \color{dgreen}70.42\% & \color{dgreen}73.59\% & \color{dgreen}71.30\% & \color{dgreen}69.54\% & \color{dgreen}71.48\% & ~\color{dgreen}71.51\% \\
            & \color{blue}BM25 & \color{blue}72.36\% & \color{blue}70.60\% & \color{blue}73.06\% & \color{blue}71.83\% & \color{blue}69.19\% & \color{blue}71.30\% & ~\color{blue}71.39\%$^\bullet$\\
            & \color{blue}CodeBERT & \color{blue}73.06\% & \color{blue}69.89\% & \color{blue}73.94\% & \color{blue}71.48\% & \color{blue}69.37\% & \color{blue}71.48\% & ~\color{blue}71.54\%$^\bullet$\\
            & \color{blue}Diversity & \color{blue}72.36\% & \color{blue}69.54\% & \color{blue}73.59\% & \color{blue}71.13\% & \color{blue}69.37\% & \color{blue}71.65\% & ~\color{blue}71.27\%$^\bullet$ \\
            & \color{blue}\textbf{CAST-F} & \color{blue}\textbf{73.24\%} & \color{blue}\textbf{70.77\%} & \color{blue}\textbf{74.12\%} & \color{blue}\textbf{72.01\%} & \color{blue}\textbf{70.07\%} & \color{blue}\textbf{71.65\%} & ~\color{blue}\textbf{71.98\%}~ \\
    \bottomrule
\end{tabular}
\end{table*}

\begin{table*}[htbp]
    \centering
    \color{blue}\caption{
        Results of various ICL strategies for 5-shot translation on AVATAR~\cite{ahmad2023avatar} of Qwen2-7B.
    }
    \begin{tabular}{lcccccccccccc}
    \toprule
        \multirow{2}{*}{\textbf{Strategy}} & \multicolumn{6}{c}{\textbf{Java to Python}} & \multicolumn{6}{c}{\textbf{Python to Java}}  \\
        \cmidrule(lr){2-7} \cmidrule(lr){8-13}
        & CA & CodeBLEU & BLEU & WNG & Syntax. & Dataflow. & CA & CodeBLEU & BLEU & WNG & Syntax. & Dataflow.  \\
    \midrule
         Random & 39.36\% & 20.06 & 1.56 & 2.29 & \textbf{36.39} & 35.40 & 46.40\% & 20.88 & 2.01 & 3.10 & 43.88 & 29.56\\
         LD & 40.56\% & 19.73 & 1.56 & 2.23 & 34.77 & 33.40 & 47.59\% & 21.30 & 1.90 & 2.90 & 42.48 & 29.96 \\
         BM25 & 46.59\% &19.85 &  1.61 & 2.26 & 36.03 & 35.50 & 44.00\% & 21.39 & 1.94 & 2.98 & 43.27 & 30.16\\
         CodeBERT & 46.59\% & 20.36 &  1.55 & 2.18 & 35.16 & \textbf{36.11} & 49.20\% & 20.88 & 1.99 & 3.14 & 44.08 & 30.30\\
         Diversity & 39.76\% & 19.72 &  1.51 & 2.14 & 34.69 & 33.70 & 44.40\% & 20.80 & 1.99 & 3.14 & 44.08 & 29.97\\
         \textbf{CAST-F} & \textbf{47.79\%} &\textbf{22.04} &  \textbf{1.73} & \textbf{2.55} & 35.79 & 33.80 & \textbf{50.40\%} & \textbf{21.55} & \textbf{2.17} & \textbf{3.38} &  \textbf{45.60} & \textbf{31.54} \\
    \bottomrule
    \end{tabular}
    \label{tab:avatar}
\end{table*}

\rqbox{How do SOTA LLMs of various scales perform in the code translation task?}
In this research question, we evaluate SOTA LLMs across various scales for code translation.
As shown in \cref{tab:baselines}, the results reveal several meaningful observations:
1)~The LLMs demonstrate impressive performance in the code translation task, highlighting their potential for practical application. Specifically, the proprietary LLMs, such as GPT-4o and GPT-4o-mini, achieve an average CA of 66.37\% and 69.72\%, respectively, indicating strong potential for LLM-based code translation.
2)~A high EM score indicates that the translated code is identical to the ground truth, often used as a metric to assess the extent of training data leakage.
Compared to learning-based methods, which ensure no data leakage during training, the EM scores of LLMs are comparable, indicating minimal data leakage. 
This shows that LLMs can effectively understand the task through prompts and produce semantically correct translated code.
3)~Most LLMs outperform learning-based models. Within the same model series, larger models, such as the Gemma2 series, perform better than smaller ones, indicating that scaling up parameters enhances the code translation capabilities of LLMs.

\findingbox{
    LLMs demonstrate remarkable performance in code translation, surpassing learning-based methods and showcasing high effectiveness and generalization. 
    Scaling up the parameters of LLMs further enhances their code translation capabilities, highlighting their strong potential for industrial applications.
}

\rqbox{How does CAST perform with SOTA LLMs compared to different ICL exemplar selection strategies?}
Referring to the results in \cref{tab:baselines}, we choose three open-source LLMs as primary baselines: Qwen2-7B, the SOTA LLM under 10B parameters; Gemma2-27B, the SOTA LLM under 40B parameters; and Qwen2.5-72B, the SOTA LLM exceeding 70B parameters. 
CAST-F is used to select exemplars for ICL, and its performance is compared against other exemplar selection strategies.
Results in \cref{tab:main_res} demonstrate the effectiveness of CAST-F across LLMs with various parameter scales in the 5-shot code translation task.
\textcolor{blue}{
We conduct t-tests comparing CAST-F with each other ICL retrieval strategies. The resulting p-values indicate that CAST-F significantly outperforms alternative strategies in most settings, with particularly strong statistical advantages observed on GPT-4o and Gemma2-27B. Methods that are highly significantly inferior to our approach ($p<0.05$) are marked with ``$\bullet$'', and those that are significantly inferior ($0.05\leq p<0.1$) are marked with ``$\circ$''.
}
There are several key observations:
1)~CAST-F consistently outperforms all other ICL exemplar selection strategies across LLMs and translation tasks.
Specifically, it surpasses the second-best strategy, Diversity, by 1.6\% for Qwen2-7B, 1.9\% for Gemma2-27B, and 1.14\% for Qwen2.5-72B. 
This shows that information coverage is a more precise and general measure for ICL, and CAST-F effectively integrates code structural knowledge into LLMs to enhance code translation. \textcolor{dgreen}{Compared with other models, since GPT-4o, as a commercial-closed model, inherently possesses strong capabilities, the performance improvement space brought by structural information injection is not significant, but CAST still maintains a certain advantage compared with baselines.}
2)~For Gemma2-27B, exemplar selection using LD, CodeBERT, and Diversity performs worse than random sampling.
This highlights the importance of effective exemplar selection strategies in ICL.
CAST-F, the only list-wise method that considers both information coverage and code structural knowledge, consistently achieves superior results compared to other strategies.
\textcolor{dgreen}{3)~AST-ED focuses more on measuring the similarity of the overall structure of the code. In contrast, CAST is based on subtree coverage, placing greater emphasis on matching local structural modules. It provides module-by-module references for translation, effectively reducing translation deviations in local logic and thereby improving the overall performance.}

\textcolor{blue}{
In real-world scenarios, the quality and structure of code repositories vary considerably. 
To more accurately assess the practical effectiveness of our method, we further evaluated its code translation performance on the AVATAR~\cite{ahmad2023avatar} dataset, which includes 249 Java samples, 250 Python samples, and 6,255 test cases.
In addition to computation accuracy, we also computed a range of metrics, including BLEU, Weighted N-Gram~(WNG) Match Score, Syntactic Match Score, Dataflow Match Score, and CodeBLEU (weighted combination of the above four parts)~\cite{codebleu}, offering a more comprehensive evaluation of translation quality. 
The experimental results, presented in \cref{tab:avatar}, reveal significant performance differences across various ICL strategies in 5-shot code translation using Qwen2-7B on AVATAR. 
Among the evaluated methods, CAST-F consistently outperforms others in both Java-to-Python and Python-to-Java translation tasks, achieving the highest computational accuracy at 47.79\% and 50.40\%, respectively, as well as the best overall CodeBLEU scores.
These results underscore CAST's strong generalization ability and its practical effectiveness in real-world code translation tasks. Specifically, CAST-F excels not only in BLEU and WNG but also in Syntactic and dataflow alignment, demonstrating its ability to capture both surface-level and structural aspects of code. In contrast, baseline strategies like Random and Diversity yield lower scores across most metrics, highlighting the importance of informed example selection in ICL-based code translation.
}

\vspace{-0.2cm}
\findingbox{
    \textcolor{blue}{CAST-F demonstrates robust performance across various LLMs, translation tasks, real-world scenarios, and multiple metrics, emphasizing its ability to capture both surface-level and structural features of code.} 
    Consequently, information coverage provides a more precise and generalized measure for ICL, with CAST-F effectively incorporating code structural knowledge into LLMs to improve code translation.    
}

\begin{figure}[!t]
    \centering
      \vspace{-0.6cm}  \includegraphics[width=0.75\linewidth]{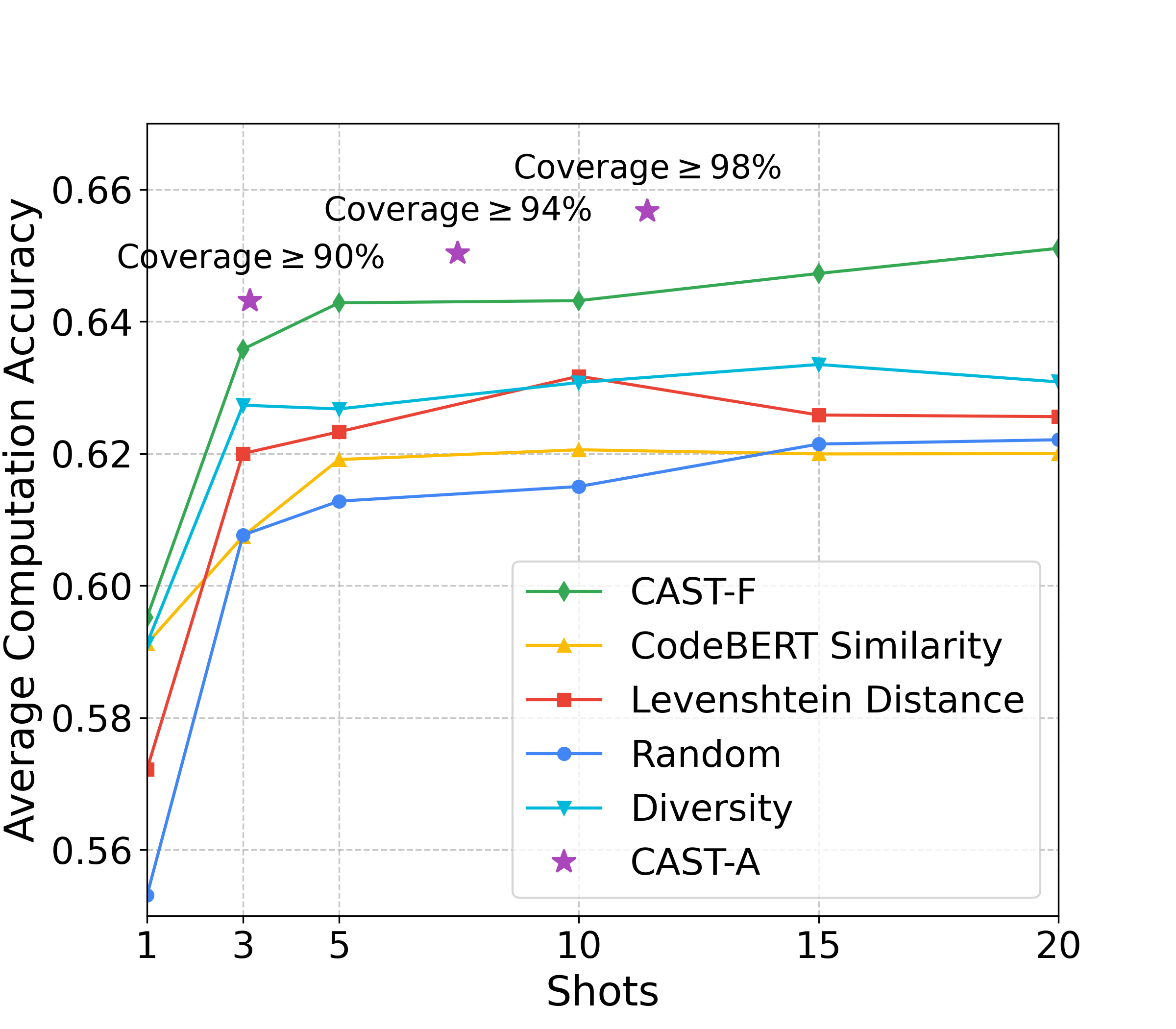}
        \caption{The performance of Qwen2-7B with various strategies and shot count.}
        \label{fig:pareto}
    \vspace{-0.5cm}
\end{figure}

\rqbox{How does ICL perform using different exemplar selection strategies as the number of exemplars increases?}
Results in \cref{tab:cmp_var_shots} and \cref{fig:pareto} show the performance of Qwen2-7B in code translation using various ICL exemplar selection strategies, with the number of exemplars increasing from $1$-shot to $20$-shots.
Several key observations arise from this analysis:
1)~CAST-F consistently achieves the highest average performance across all strategies, with an accuracy of 65.11\% at $20$-shots. 
2)~Increasing the number of exemplars generally improves performance across most strategies. CAST-F demonstrates superior scalability, particularly at higher shot counts, underscoring its robustness in improving code translation.
3)~Compared to baselines, CAST-F consistently delivers incremental gains by leveraging information coverage and code structural knowledge.
4)~The Diversity strategy ranks second, performing well by considering both point-wise similarity and diversity, but it lacks the ability to account for code structural knowledge or perform selection in a compositional manner.
5)~CAST-F achieves the highest accuracy across challenging language pairs, such as Python to Java~(58.10\%) and Java to Python~(69.37\%), demonstrating its effectiveness in translating between syntactically diverse languages.
6)~A law of diminishing marginal utility is evident in \cref{tab:cmp_var_shots} and \cref{fig:pareto} as the number of exemplars increases. While performance improves significantly with the first few exemplars, additional gains become minimal beyond $5$-shots due to redundancy or information saturation. However, Qwen2-7B can remit this issue effectively by considering information coverage as a more precise and general measure.

\findingbox{
    Despite diminishing marginal utility with more exemplars in all ICL strategies, CAST-F shows superior performance and scalability. It achieves the highest accuracy at higher shot counts by considering information coverage and code structure.
}

\begin{table*}[!t]
    \centering
    \setlength\tabcolsep{1.1em}
    \caption{
        Performance of Qwen2-7B on code translation with various ICL exemplar selection strategies and numbers of exemplars.
    }
    \label{tab:cmp_var_shots}
    \begin{tabular}{lcccccccc}
        \toprule
             \textbf{Strategy} & \textbf{\#Shots} & \textbf{C++ to Py} & \textbf{Py to C++} & \textbf{Java to C++} & \textbf{C++ to Java} & \textbf{Py to Java} & \textbf{Java to Py} & \textbf{Average} \\ 
        \midrule
            Fixed & 1 & 58.27\% & 50.70\% & 60.74\% & 50.88\% & 20.07\% & 63.20\% & 50.64\% \\ 
        \midrule
            \multirow{6}{*}{Random} 
            & 1 &
                60.39\% & 54.93\% & 66.90\% & 52.11\% & 29.93\% & 67.60\% & 55.31\% \\
            & 3 &
                63.02\% & 57.02\% & 65.14\% & 61.90\% & 49.74\% & 67.78\% & 60.77\% \\
            & 5 &
                63.38\% & 59.31\% & 66.55\% & 61.27\% & 49.74\% & 67.43\% & 61.28\% \\
            & 10 &
                63.73\% & 57.57\% & 66.55\% & 63.73\% & 49.65\% & 67.78\% & 61.50\% \\
            & 15 &
                63.20\% & 59.86\% & 66.73\% & 64.96\% & 50.70\% & 67.42\% & 62.15\% \\
            & 20 &
                63.56\% & 59.15\% & 66.02\% & 64.08\% & 51.94\% & 68.49\% & 62.21\% \\
        \midrule
            \multirow{6}{*}{LD}
            & 1 &
                59.51\% & 59.51\% & 67.43\% & 54.57\% & 34.51\% & 67.78\% & 57.22\% \\
            & 3 &
                61.44\% & 60.92\% & 67.78\% & 63.38\% & 51.06\% & 67.43\% & 62.00\% \\
            & 5 &
                64.96\% & 57.69\% & 67.96\% & 63.73\% & 52.03\% & 67.61\% & 62.33\% \\
            & 10 &
                64.26\% & 60.73\% & 67.43\% & 65.67\% & 52.46\% & 68.49\% & 63.17\% \\
            & 15 &
                63.73\% & 59.15\% & 66.90\% & 65.49\% & 52.99\% & 67.25\% & 62.59\% \\
            & 20 &
                64.44\% & 59.15\% & 67.43\% & 64.43\% & 52.64\% & 67.25\% & 62.56\% \\
        \midrule
           \multirow{6}{*}{\color{blue}BM25} 
            & \color{blue}1 & \color{blue}63.20\% & \color{blue}57.04\% & \color{blue}66.37\% & \color{blue}57.04\% & \color{blue}40.32\% & \color{blue}67.61\% & \color{blue}58.60\% \\
            & \color{blue}3 & \color{blue}64.26\% & \color{blue}58.45\% & \color{blue}67.78\% & \color{blue}62.68\% & \color{blue}49.65\% & \color{blue}67.78\% & \color{blue}61.77\% \\
            & \color{blue}5 & \color{blue}64.61\% & \color{blue}58.98\% & \color{blue}67.25\% & \color{blue}63.73\% & \color{blue}50.18\% & \color{blue}67.78\% & \color{blue}62.09\% \\
            & \color{blue}10 & \color{blue}64.44\% & \color{blue}59.15\% & \color{blue}67.25\% & \color{blue}63.56\% & \color{blue}53.17\% & \color{blue}68.13\% & \color{blue}62.62\% \\
            & \color{blue}15 & \color{blue}65.14\% & \color{blue}60.56\% & \color{blue}67.08\% & \color{blue}64.08\% & \color{blue}52.46\% & \color{blue}68.13\% & \color{blue}62.91\% \\
            & \color{blue}20 & \color{blue}65.84\% & \color{blue}60.74\% & \color{blue}67.78\% & \color{blue}65.32\% & \color{blue}53.35\% & \color{blue}68.31\% & \color{blue}63.56\% \\
        \midrule
            \multirow{6}{*}{CodeBERT}
            & 1 &
                64.61\% & 56.69\% & 64.61\% & 54.92\% & 46.13\% & 67.78\% & 59.12\% \\
            & 3 &
                63.91\% & 57.39\% & 66.73\% & 62.15\% & 47.05\% & 67.25\% & 60.75\% \\
            & 5 &
                64.79\% & 57.02\% & 66.02\% & 63.91\% & 53.35\% & 66.37\% & 61.91\% \\
            & 10 &
                64.79\% & 56.16\% & 65.49\% & 64.08\% & 53.17\% & 68.66\% & 62.06\% \\
            & 15 &
                65.32\% & 55.96\% & 65.32\% & 64.43\% & 53.17\% & 67.78\% & 62.00\% \\
            & 20 &
                64.44\% & 57.75\% & 65.49\% & 61.62\% & 54.23\% & 68.49\% & 62.00\% \\
        \midrule
        \multirow{6}{*}{Diversity}
            & 1 &
                64.61\% & 56.69\% & 64.61\% & 54.92\% & 46.13\% & 67.78\% & 59.12\% \\
            & 3 & 64.08\%
                 & 58.63\% & 67.42\% & 64.26\% & 54.05\% & 67.96\% & 62.73\%  \\
            & 5 & 64.96\%
                 & 59.51\% & 67.42\% & 65.15\% & 51.41\% & 67.61\% & 62.68\% \\
            & 10 & 64.79\%
                 & 59.33\% & 67.42\% & 65.49\% & 53.12\% & 68.31\% & 63.08\%  \\
            & 15 & 64.96\%
                 & 60.39\% & 67.42\% & 65.85\% & 53.52\% & 67.96\% & 63.35\% \\
            & 20 & 65.14\%
                 & 60.74\% & 66.02\% & 65.49\% & 52.29\% & 68.84\% & 63.09\% \\
        \midrule
            \multirow{6}{*}{\textbf{CAST-F}} 
            & 1 &
                64.08\% & 61.09\% & 67.96\% & 59.96\% & 35.72\% & 68.13\% & 59.52\% \\
            & 3 &
                64.61\% & 61.09\% & 68.31\% & 64.44\% & 54.74\% & 68.13\% & 63.58\% \\
            & 5 &
                65.32\% & 61.62\% & 68.84\% & 65.15\% & 55.77\% & 69.01\% & 64.28\% \\
            & 10 &
                65.14\% & 60.92\% & 68.49\% & 66.19\% & 56.16\% & 69.01\% & 64.32\% \\
            & 15 &
                65.84\% & 61.09\% & 68.66\% & 66.37\% & 57.22\% & 69.19\% & 64.73\% \\
            & 20 &
                66.02\% & 61.27\% & 69.37\% & 66.55\% & \textbf{58.10\%} & \textbf{69.37\%} & 65.11\% \\
        \midrule
           Average \#Shots & \multirow{2}{*}{11.43} & 
               15.65 & 8.08 & 13.98 & 9.57 & 7.37 & 13.94 & 11.43 \\
           \textbf{CAST-A~($\bm{\geq 98\%}$)} &  & 
               \textbf{67.25\%} & \textbf{61.80\%} & \textbf{71.13\%} & \textbf{68.31\%} & 56.16\% & \textbf{69.37\%} & \textbf{65.67\%} \\
        \bottomrule
    \end{tabular}
\end{table*}

\begin{figure*}[!t]
    \vspace{-0.5cm}
    \centering
    \subfloat[C++ to Java]{%
        \includegraphics[width=0.245\textwidth]{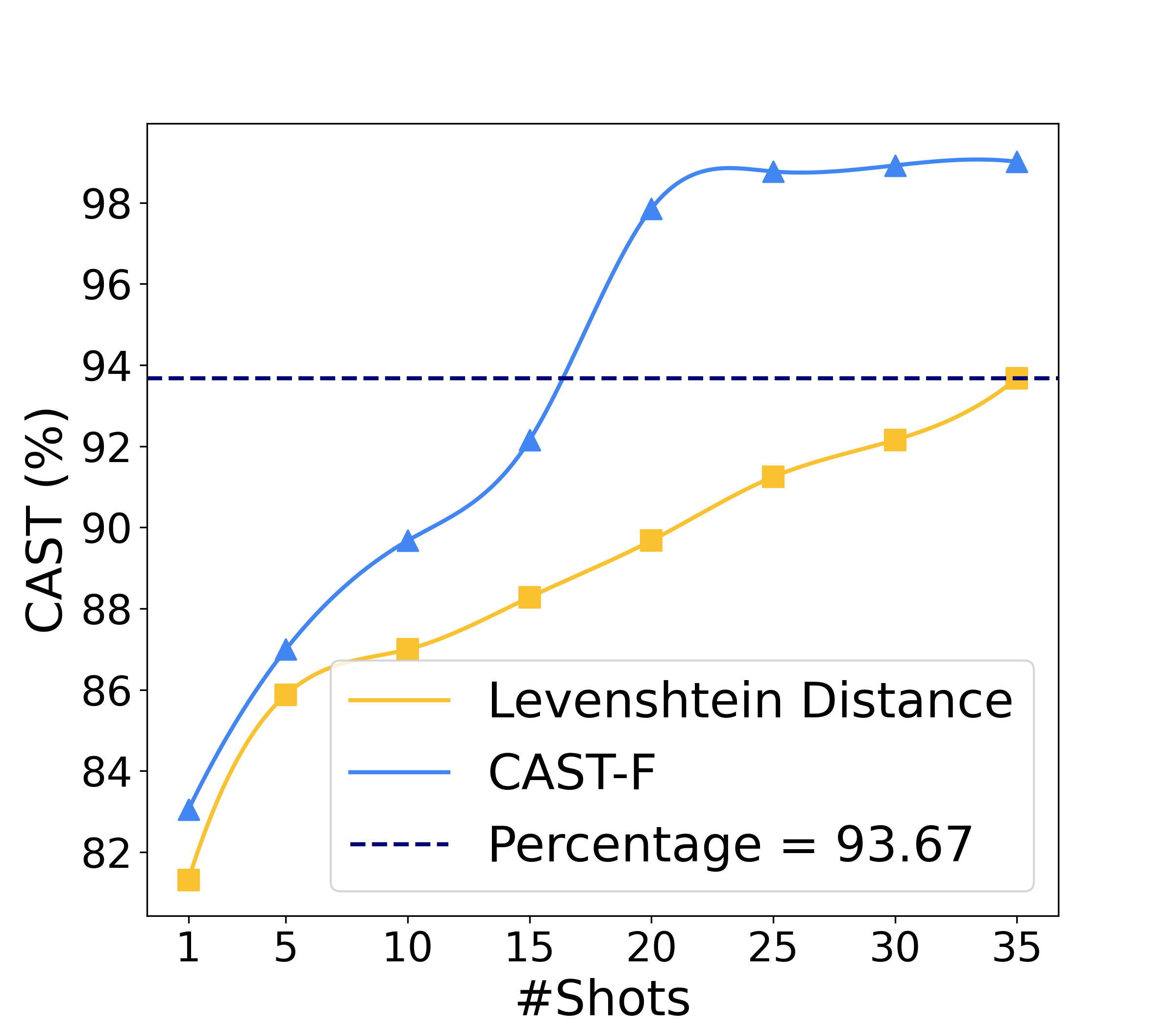}%
        \label{fig:javacpp}%
    }\hfill
    \subfloat[Python to C++]{%
        \includegraphics[width=0.245\textwidth]{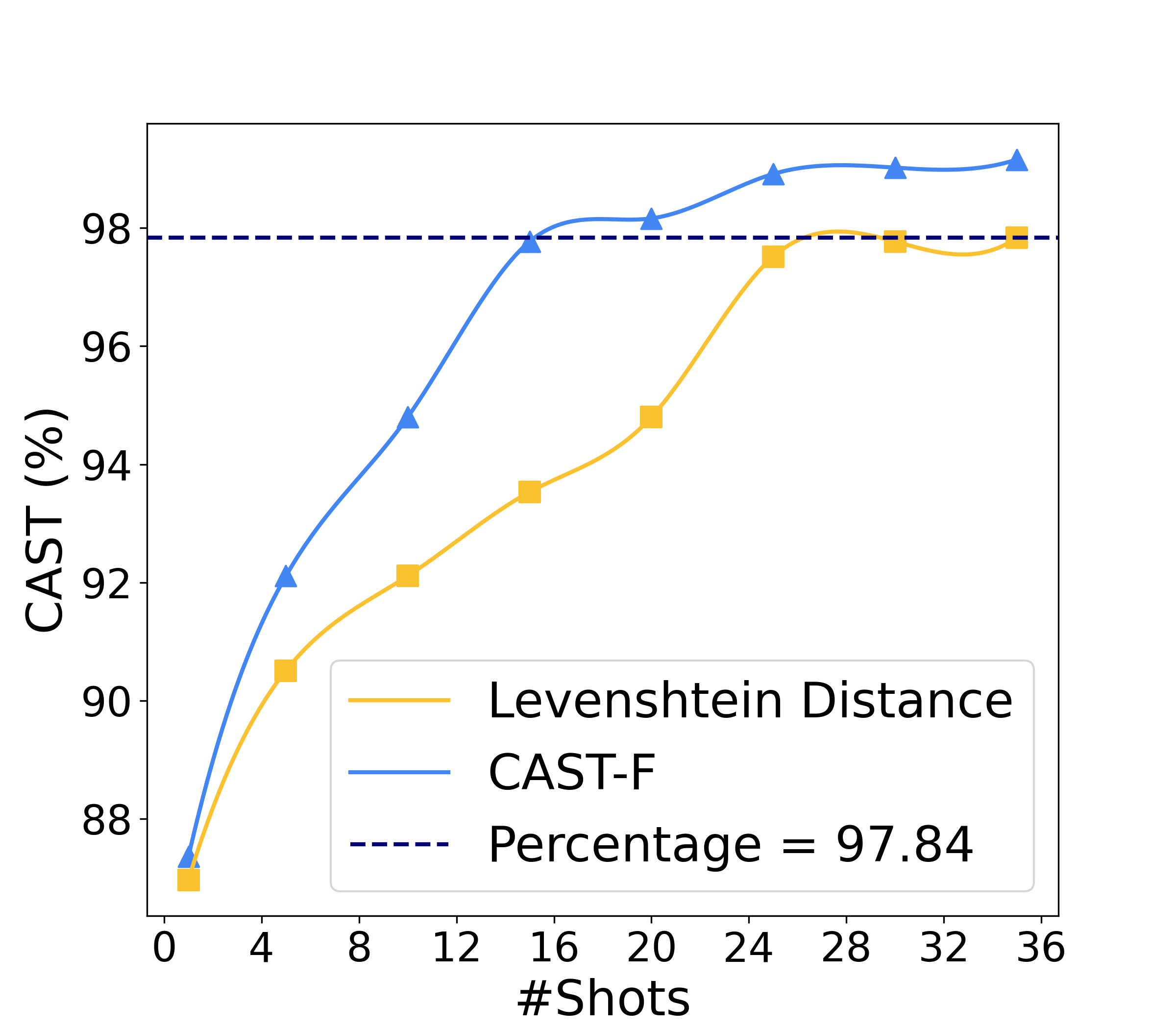}%
        \label{fig:cppjava}%
    }\hfill
    \subfloat[Python to Java]{%
        \includegraphics[width=0.245\textwidth]{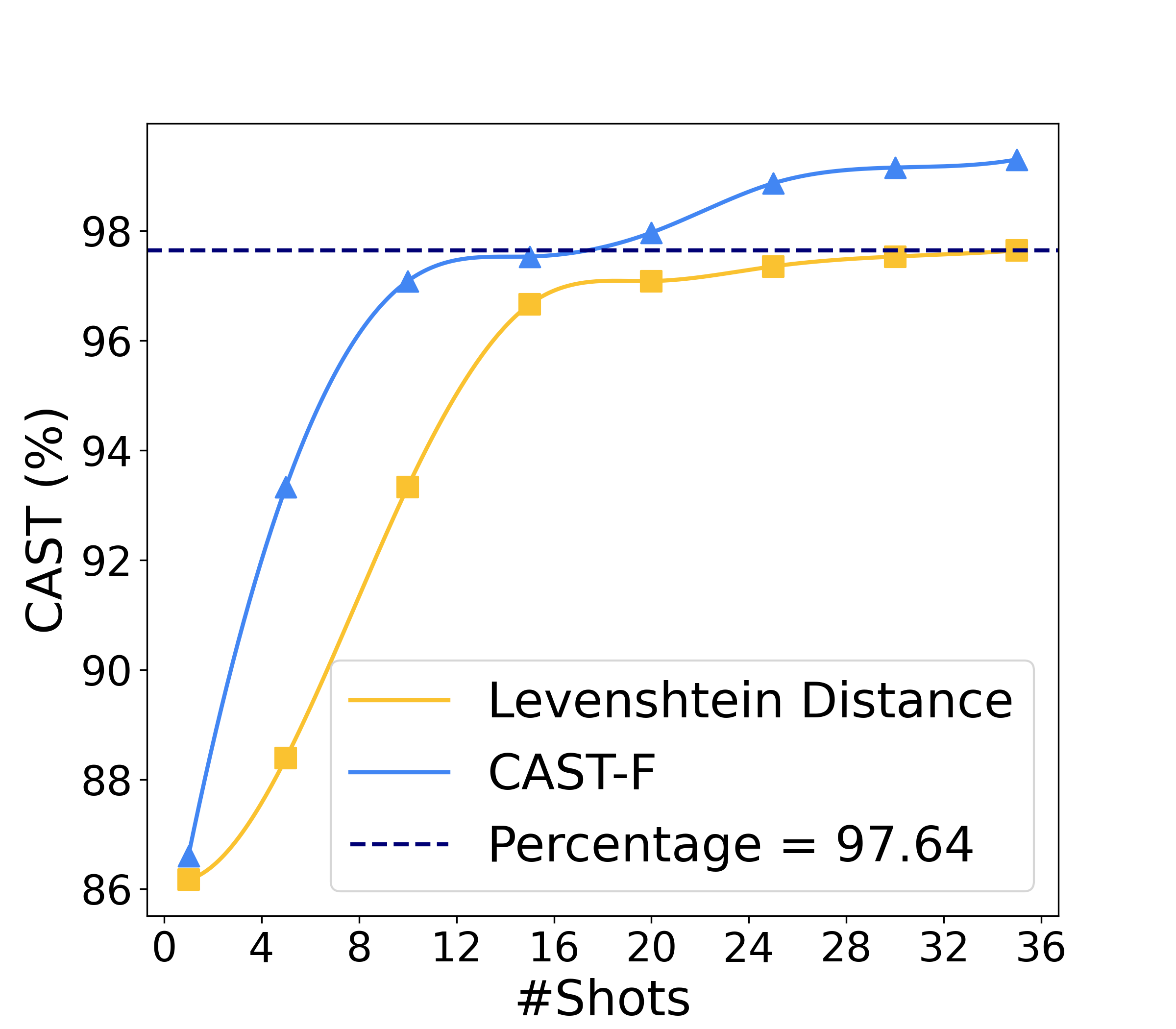}%
        \label{fig:pythoncpp}%
    }\hfill
    \subfloat[C++ to Python]{%
        \includegraphics[width=0.245\textwidth]{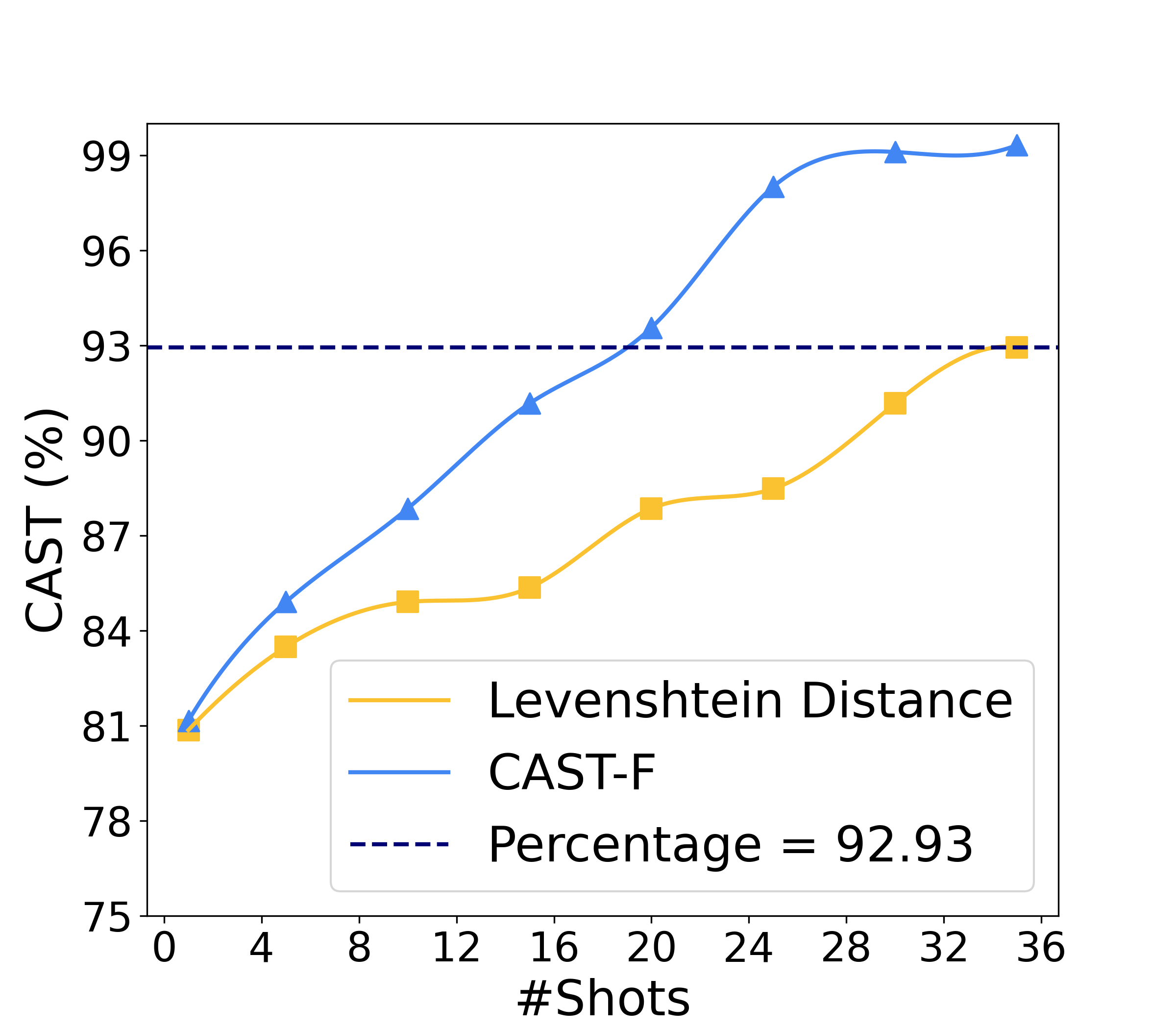}%
        \label{fig:cpppython}%
    }
    \caption{Comparison of the average coverage of abstract syntax tree~(CAST) between the CAST-F and Levenshtein distance~(LD) strategies.}
    \label{fig:coverage}
    \vspace{-0.5cm}
\end{figure*}

\rqbox{How does CAST-A perform when dynamically determining the selection size?}
Results of CAST-A, as presented in \cref{tab:cmp_var_shots} and \cref{fig:pareto}, reveal several key observations: 
1)~CAST-A achieves competitive performance compared to fixed-size exemplar selection strategies with fewer exemplars. 
It dynamically adjusts the selection size based on a CAST threshold~(e.g., CAST$\geq$ 98\%), surpassing all baselines at $20$-shots while using an average of only 11.43 exemplars.
2)~Compared to fixed-shot strategies, CAST-A mitigates the issue of diminishing returns observed in static exemplar selection by dynamically adjusting the selection size to balance information saturation and redundancy.  
3)~CAST-A demonstrates superior adaptability to varying input complexity.
As shown in \cref{fig:pareto}, CAST-F forms the Pareto frontier between input complexity~(exemplar count) and performance~(average CA) across all fixed-shot strategies. 
Surprisingly, CAST-A with different thresholds occupies the front of the Pareto frontier of fixed-strategies, showcasing the high effectiveness, efficiency, and robustness of both CAST-F and CAST-A.

\findingbox{
    CAST-A effectively adapts the selection size based on the CAST threshold, outperforming fixed-size strategies while using fewer exemplars. 
    This dynamic adjustment mitigates diminishing returns and improves scalability, showcasing its robustness in handling varying input complexities.
}

\rqbox{How does the CAST convergence of CAST-F compare with that of similarity-only strategy~(e.g., LD)?}
CAST serves as a surrogate measure of information, indicating the ratio of necessary knowledge gained from exemplars during code translation.
As shown in \cref{fig:coverage}, CAST is computed from exemplars retrieved by CAST-F and LD across varying shot counts, revealing several key observations:
1)~CAST-F consistently converges faster than LD as the number of shots increases, demonstrating greater efficiency in acquiring necessary knowledge from exemplars for translation tasks.
2)~CAST-F achieves a higher final CAST ratio than LD across all translation tasks, showcasing its robustness in optimizing exemplar selection. This emphasizes the advantage of CAST-F in handling complex translations.

\findingbox{
    CAST-F accelerates the acquisition of necessary knowledge and ensures higher final gains across all tasks. These advantages demonstrate its effectiveness in optimizing exemplar selection for diverse and complex code translation scenarios.
}

\begin{table*}[!t]
    \centering
    \setlength\tabcolsep{0.88em}
    \caption{
            Results of the gain of CAST-F among different sizes of LLMs, consisting of 2B, 9B, and 27B of Gemma2.}
    \label{tab:params}
    \begin{tabular}{llcccccccccc}
        \toprule
            \multirow{2.5}{*}{\textbf{Models}} & \multirow{2.5}{*}{\textbf{Strategy}} &
            \multicolumn{2}{c}{\textbf{C++ to Python}} & \multicolumn{2}{c}{\textbf{Python to C++}} & \multicolumn{2}{c}{\textbf{Java to C++}} & \multicolumn{2}{c}{\textbf{C++ to Java}} & \multicolumn{2}{c}{\textbf{Average}} \\    
            
            \cmidrule(lr){3-4} \cmidrule(lr){5-6} \cmidrule(lr){7-8} \cmidrule(lr){9-10} \cmidrule{11-12}
            
            \multicolumn{2}{c}{} & 
            \textbf{CA} & \textbf{EM} & 
            \textbf{CA} & \textbf{EM} &
            \textbf{CA} & \textbf{EM} &
            \textbf{CA} & \textbf{EM} &
            \textbf{CA} & \textbf{EM} \\
        \midrule
            \multirow{3}{*}{Gemma2-2B} 
            & Fixed & 47.36\% & 6.51\% & 39.61\% & 2.64\% & 54.05\% & 12.15\% & 22.18\% & 5.28\% & 40.80\% & 6.65\% \\
            & CAST-F & 51.06\% & 8.10\% & 48.77\% & 3.17\% & 63.56\% & 17.43\% & 30.81\% & 10.39\% & 48.55\% &	9.77\% \\ 
            \cmidrule(lr){2-12}
            & \emph{Improv.} & 7.81\%& 	24.42\%	& \textbf{23.13\%}& 	20.08\%	& \textbf{17.59\%}& 	\textbf{43.46\%}	& 38.91\%	& \textbf{96.78\%}& 	21.86\%	& 46.18\% \\
        \midrule
            \multirow{3}{*}{Gemma2-9B} 
            & Fixed &
                60.39\% & 8.45\% & 59.26\% & 4.75\% & 65.61\%  & 31.34\%  & 20.81\%  &  23.06\% &  51.52\% & 16.90\% \\ 
            & CAST-F  & 60.74\% & 14.08\% & 64.08\% & 9.51\% & 69.19\% & 36.27\% & 65.14\% & 32.22\% & 64.79\% &	23.02\%\\ \cmidrule(lr){2-12}
            & \emph{Improv.} & 0.58\%	& \textbf{66.63\%}	&8.13\%	&\textbf{100.21\%}	&5.46\%	&15.73\%	&\textbf{213.02\%} &	39.72\%	&\textbf{56.80\%}	&\textbf{55.57\%} \\
        \midrule
            \multirow{3}{*}{Gemma2-27B} 
            & Fixed &
                 63.38\% & 8.45\% & 63.91\% & 5.33\% & 68.49\% & 24.96\% & 68.31\% & 21.83\% & 66.02\%	& 15.14\% \\ 
            & CAST-F & 70.60\% & 11.62\% & 68.66\% & 6.87\% & 71.13\% & 34.86\% & 72.49\% & 28.87\% & 70.72\%	& 20.56\% \\ \cmidrule(lr){2-12}
            & \emph{Improv.} & \textbf{11.39\%} &	37.51\% &	7.43\% &	28.89\% &	3.85\% &	39.66\% &	6.12\% &	32.25\% &	7.20\% &	34.58\% \\
        \bottomrule
    \end{tabular}
\end{table*}

\rqbox{
   Does scaling up the model size automatically lead to sufficient code structural knowledge?
}
As shown in \cref{tab:main_res}, although translation performance improves with larger models, the improvement from explicitly incorporating the code structural knowledge by CAST-F is much more substantial.
Additionally, for a fairer comparison, we evaluate CAST-F with $5$-shots on Gemma2-(2B/9B/27B), all of which are from the same family. 
The results, shown in \cref{tab:params}, indicate that the improvement from CAST-F is not significantly related to model scale.
This confirms that the additional code structural knowledge from CAST-F is independent of model scale.
Therefore, simply scaling up the model or training data does not result in the sufficient emergence of code structural knowledge, highlighting the necessity of explicitly incorporating code syntax.

\findingbox{
    Scaling up model size or training data does not lead to the sufficient emergence of code structural knowledge, underscoring the necessity of explicitly considering code syntactic structure in LLMs.
}

\begin{figure*}[!t]
    \centering
    \begin{minipage}[t]{0.315\linewidth}
        \centering
        \includegraphics[width=0.8\linewidth]{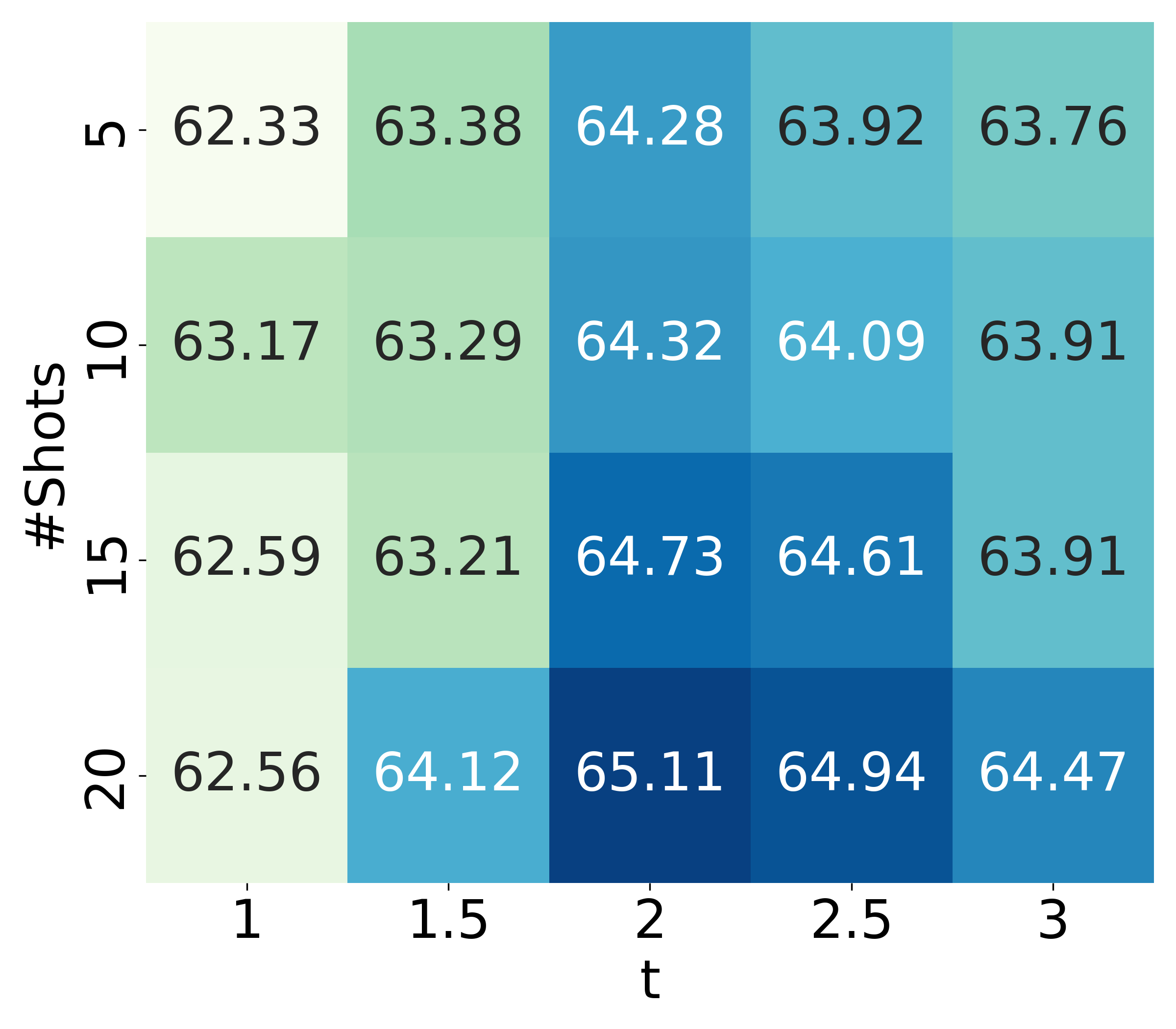}
        \caption{
            Sensitivity analysis of hyper-parameters in the average CA of Qwen2-7B for the selection size $k$ and the pre-recalled candidate size $\lfloor t \cdot k \rfloor$.
        }
        \label{fig:heat}
    \end{minipage}
    \hfill
    \begin{minipage}[t]{0.65\linewidth}
        \centering
        \includegraphics[width=\linewidth]{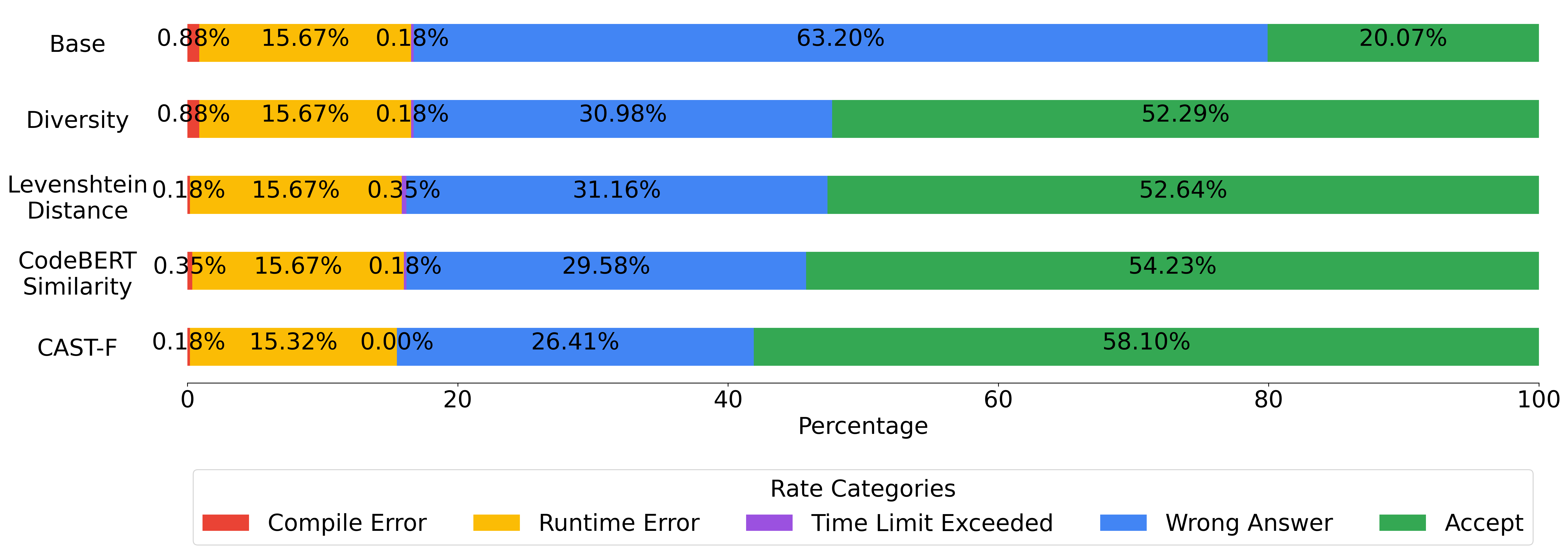}
        \caption{
            The distribution of result types in Python-to-Java 20-shots translation by Qwen2-7B across various exemplar selection strategies in ICL.
            Outcomes are categorized as Compile Error, Runtime Error, Time Limit Exceeded, Wrong Answer, and Accept.
        }
        \label{fig:fail_type}
    \end{minipage}
\end{figure*}

\rqbox{What is the hyperparameter sensitivity?}
In CAST-F, there is only one hyperparameter, 
i.e., the factor $t$ for adjusting the size of lexical pre-recall~($\lfloor t \times k \rfloor$).
We evaluate CAST-F across varying values of the factor $t$ and the selection size $k$.
\cref{fig:heat} shows the results, which indicate that regardless of the selection size $k$, $t=2$ yields the best performance, and performance remains stable across all values of $k$ for varying $t$.
For CAST-A, there is an additional threshold $\tau$. 
Results in \cref{fig:pareto} indicate that CAST-A achieves Pareto optimality when $\tau$ varies between 90\% and 98\%, demonstrating stable and effective performance across different thresholds.

\findingbox{
    Both CAST-F and CAST-A demonstrate strong robustness and stability across varying hyperparameter settings.
}

\rqbox{What is the impact of our method on failure frequencies?}
This study evaluates how exemplar selection strategies influence test code failures across various input cases.
Results in \cref{fig:fail_type} highlight key observations:
1)~CAST-F achieves the lowest proportion of ``Wrong Answer'' outcomes (26.41\%) and the highest ``Accept'' rate (58.10\%), surpassing Levenshtein Distance (52.64\%) and CodeBERT Similarity (54.23\%). 
This underscores its superior effectiveness in selecting exemplars to enhance translation accuracy.
2)~ICL yields limited improvements for ``Compile Error'' and ``Runtime Error.'' These limitations likely stem from cross-language differences and the lack of comprehensive third-party library knowledge, which hinder LLMs' ability to resolve such errors.

\findingbox{
    CAST-F effectively reduces failure frequencies, achieving higher accuracy and robustness compared to baseline strategies, particularly in minimizing ``Wrong Answer'' outcomes and maximizing ``Accept'' rates.
}



\begin{table*}[htbp]
    \centering
    \setlength\tabcolsep{0.84em}
    \color{blue}\caption{
        Results of Qwen2-7B for 5-shot Code summarization upon different ICL strategies.
    }
    \begin{tabular}{lcccccccccccc}
    \toprule
        \multirow{2}{*}{\textbf{Strategy}} & \multicolumn{2}{c}{\textbf{Java}} & \multicolumn{2}{c}{\textbf{Python}} & \multicolumn{2}{c}{\textbf{PHP}} & \multicolumn{2}{c}{\textbf{Go}} & \multicolumn{2}{c}{\textbf{Javascript}} & \multicolumn{2}{c}{\textbf{Ruby}} \\
        \cmidrule(lr){2-3} \cmidrule(lr){4-5} \cmidrule(lr){6-7} \cmidrule(lr){8-9} \cmidrule(lr){10-11} \cmidrule(lr){12-13}
         & BLEU &  Rouge-L & BLEU &  Rouge-L & BLEU &  Rouge-L & BLEU &  Rouge-L & BLEU &  Rouge-L & BLEU &  Rouge-L \\
    \midrule
        Random & 2.25 & 9.01 & 5.29 & 14.80 & 12.30 & 10.78 & 4.60 & 14.93 & 2.33 & 5.45 & 5.02 & 13.13\\
        BM25 & 2.27 & 9.60 & 5.92 & 14.74 & \textbf{17.65} & 16.55 & 4.48 & 14.24 & 2.86 & 6.03 & 5.49 & 12.50 \\
        LD & 2.06 & 9.39 & 4.47 & 13.45 & 16.27 & 16.14 & 5.43 & 15.63 & 3.11 & 7.83 & 4.82 & 13.00 \\
        CodeBERT & \textbf{2.43} & 9.56 & 5.92 & 15.07 & 17.41 & 15.05 & 4.34 & 13.35 & 2.80 & 5.87 & 5.99 & 13.14 \\
        Diversity & 2.30 & 9.39 & 5.23 & 13.00 & 14.52 & 13.40 & 3.88 & 12.80 & 2.43 & 5.86 & 5.25 & 13.34\\
        \textbf{CAST-F} & 2.42 & \textbf{9.90} & \textbf{6.73} & \textbf{16.57} & 17.58 & \textbf{18.65} & \textbf{5.52} & \textbf{17.68} & \textbf{4.81} & \textbf{9.12} & \textbf{6.71} & \textbf{14.47} \\
    \bottomrule
    \end{tabular}
    \label{tab:codesum}
\end{table*}

\textcolor{blue}{
\rqbox{Could CAST be extended to support programming languages and tasks beyond code translation?}
To showcase the broader applicability of CAST beyond code translation, we conduct an additional experiment on code summarization. 
Code summarization generates natural language descriptions from source code snippets, a crucial task in software maintenance and comprehension. 
We use the widely used dataset provided by the CodeXGLUE~\cite{CodeXGLUE} team for this task.
We evaluate the performance of our method, using the widely adopted Qwen2-7B model across six programming languages: Java, Python, PHP, Go, JavaScript, and Ruby.
The evaluation metrics include BLEU~\cite{papineni2002bleu} and Rouge-L scores, assessing the quality of generated summaries against reference descriptions. 
Rouge-L computes an F-score based on the longest common subsequence, comparing the similarity between two texts.
}

\textcolor{blue}{
\cref{tab:codesum} presents results comparing CAST-F with baseline exemplar selection strategies. 
CAST-F achieves the highest BLEU and Rouge-L scores across nearly all languages and metrics, demonstrating its superior ability to select exemplars that enhance the model's summarization quality. 
This consistent improvement underscores that CAST’s structural coverage principle captures critical code characteristics essential for generating accurate and informative summaries. These findings emphasize CAST’s generality and robustness as an exemplar selection metric.
By capturing structural coverage effectively, CAST facilitates improved task-specific knowledge transfer for LLMs in code-related tasks.
}

\findingbox{
\textcolor{blue}{The experiment on code summarization broadens the scope of our approach beyond code translation, establishing CAST as a versatile tool for enhancing code retrieval and generation tasks in various software engineering scenarios.}
}

\textcolor{blue}{
\rqbox{How do different database sizes and noisy data affect the method's performance?}
We evaluated the impact of database size variations and noise injection on CAST performance using the TransCoder dataset Java-Python task under Qwen-7B.
To assess database size effects, we randomly sampled subsets at 75\%, 50\%, and 25\% of the full database as ICL demonstration exemplars, simulating scenarios with limited available data. 
For noise injection, we implemented three perturbation types to simulate real-world code imperfections:
\begin{itemize}
    \item \textit{Syntax Error}: Random deletion or replacement of punctuation marks~(semicolons, parentheses, quotes), introducing minor cross-language syntax errors.
    \item \textit{Identifier Rename}: Variable name obfuscation or random replacement, breaking identifier consistency and simulating naming errors.
    \item \textit{Statement Shuffle}: Random reordering of consecutive statements within code snippets, mimicking structural disorder in semantically related blocks.
\end{itemize}
}

\begin{figure}[!t]
    \centering
    \subfloat[Database Size Impact]{%
        \includegraphics[width=0.48\linewidth]{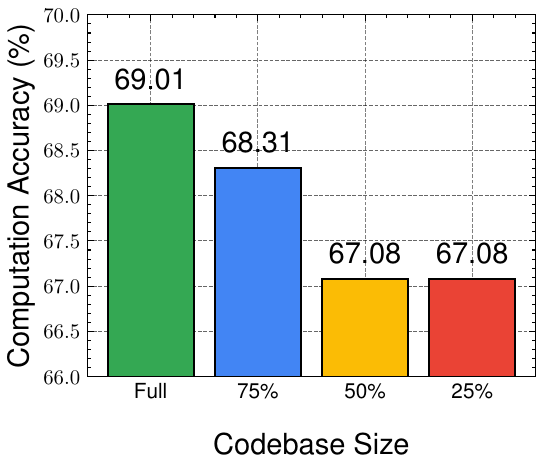}%
        \label{fig:db_size}%
    }\hfill
    \subfloat[Database Noise Impact]{%
        \includegraphics[width=0.48\linewidth]{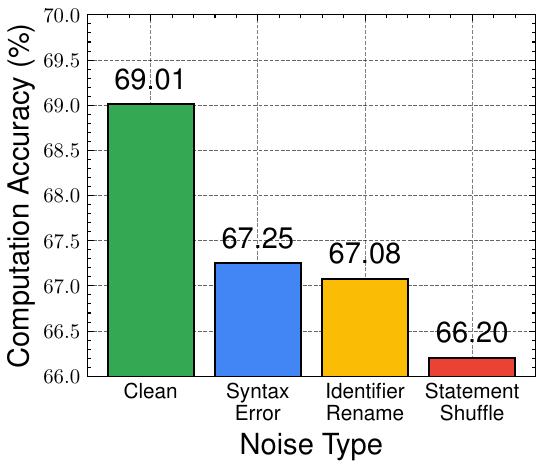}%
        \label{fig:db_noise}%
    }
    \caption{\textcolor{blue}{
        Impact of database characteristics on computation accuracy. 
        \cref{fig:db_size} Database size effect with random sampling.
        \cref{fig:db_noise} Performance under different noisy database.
    }}
\end{figure}

\textcolor{blue}{
As shown in \cref{fig:db_size}, reducing the database size for exemplar sampling leads to gradual accuracy degradation. Performance decreases from 69.01\% (full database) to 67.08\% (25\% sampling), indicating that while CAST maintains robustness under reduced exemplar availability, performance diminishes as database information decreases.
\cref{fig:db_noise} demonstrates performance degradation under different noise types. Accuracy drops from 69.01\% (no noise) to 66.20\% (statement shuffle), with syntax errors and identifier renaming showing intermediate impact (67.25\% and 67.08\% respectively). These results indicate CAST's relative resilience to syntax and naming perturbations but higher sensitivity to structural disruptions.
}

\findingbox{
\textcolor{blue}{CAST performance degrades with reduced database size (2.93\% drop at 25\% sampling) and structural noise (2.81\% loss with statement shuffle). Adequate database scale and code quality are essential for optimal performance.}
}

\begin{figure*}[!t]
    \centering
    \includegraphics[width=\linewidth]{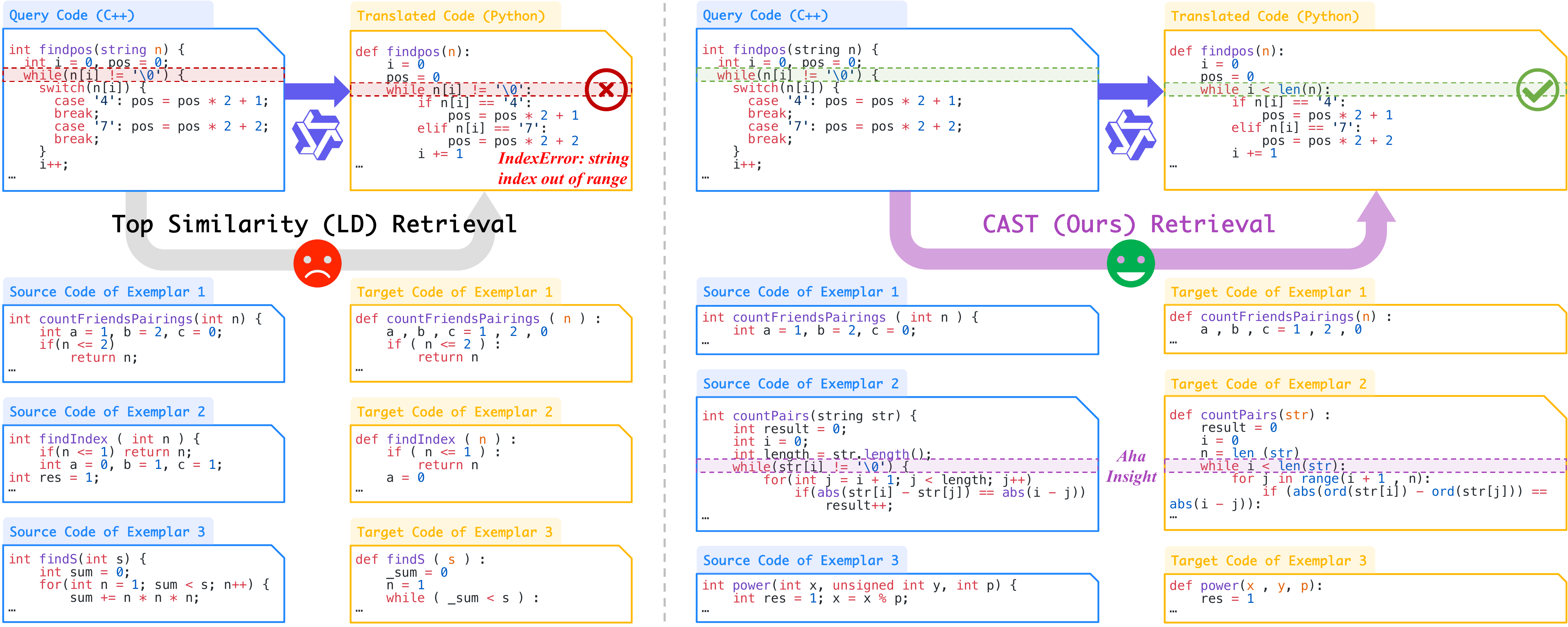}
    \caption{\textcolor{blue}{
        A case study~(qualitative analysis) of code translation comparing similarity-only method~(i.e., LD) and our CAST-F, based on Qwen2-7B.
    }}
    \label{fig:casestudy}
\end{figure*}

{\color{blue}
\section{Case Study}
To understand why traditional similarity-based retrieval fails to provide optimal exemplars for code translation, we conduct a case study examining how different retrieval strategies affect translation quality.
The fundamental limitation of similarity-only approaches is their tendency to retrieve exemplars with high similarity while overlooking sufficient knowledge required for the current translation task.
\cref{fig:casestudy} shows this through a C++ to Python translation task about \texttt{findpos} function that processes string-based position calculations.

The similarity-only retrieval~(LD) operates purely on lexical similarity, leading to the selection of highly correlated but redundant exemplars that cannot provide sufficient knowledge for the current translation task.
The case shows that the translated code fails in string boundary checking, resulting in an \texttt{IndexError: string index out of range}, where correctly translating this construct requires knowledge of the difference between C++ null-terminator-based string handling (\lstinline[style=customcpp]|'\0'|) and Python's length-based indexing (\lstinline[style=customcpp]|len()|). 
However, the retrieved exemplars fail to provide this critical cross-language knowledge for properly handling the string termination construct \lstinline[style=customcpp]|while (n[i] != '\0')|.

In contrast, CAST retrieval functions as a surrogate objective for information coverage across syntactic structures, actively seeking exemplars that provide the knowledge required for the current translation task.
By retrieving exemplars that collectively cover diverse syntactic structural patterns, CAST enables the model to discover the ``aha insight'' that Python requires length-based indexing (\lstinline[style=customcpp]|while i < len(n)|).
This demonstrates how CAST's information coverage principle guides the model toward semantically correct translations by ensuring comprehensive knowledge representation in the retrieved exemplar set.
The success stems from CAST's ability to prioritize knowledge completeness over superficial similarity, leading to more robust code translation.
}

\section{Threat Analysis}
\label{threat}
Our results are interpreted with three threats to validity.
\begin{itemize}
    \item The internal validity threat comes from potential LLM data leakage due to possible training/testing set overlaps, but prior work~\cite{yang2024exploring} shows this threat is limited.
    \item The first external validity threat is in the implementation of the compared techniques. To reduce it, we select state-of-the-art pre-trained models and reuse their implementations and weights.
    We believe this threat is minimal.
    \item Another external validity threat is the quality of the dataset. It was released by~\cite{transcoder1} and is widely used due to multi-lingual samples and self-contained test suites~\cite{transcoder2,transcoder3,yang2024exploring}. We use the cleaned dataset from~\cite{yang2024exploring}, where four authors cross-checked their cleaned results in pairs to reach a consensus for each group of parallel functions.
    \item The construct threat lies in the evaluation metrics. To reduce this threat, we adopt CA and EM to evaluate the code translation performance of different approaches from both lexical and semantic correctness as most of the relevant papers~\cite{transcoder1,transcoder2,transcoder3,yang2024exploring}. 
\end{itemize}

\section{Related Works}

\subsection{Code translation}
Automated code translation aids in the migration of codebases across programming languages, significantly enhancing productivity in software development.
Early rule-based translation methods have been extensively explored~\cite{rule1,rule2,rule3,rule4}.
Furthermore, phrase-based statistical machine translation techniques have shown notable advances~\cite{KaraivanovRV14,NguyenNN15,NguyenNN13,NguyenNN16a,AllamanisBDS18,swt}.
Due to the success of deep learning, deep code translation methods have extended the SOTA further~\cite{LampleCDR18,babel,pt5,must,pscpt,du2024joint}.

Prior research has shown that, unlike the grammatical structure of natural language, 
the syntactic structure of source code is stricter and more intricate, directly reflecting the program's behavior~\cite{tree2tree,sem4,sem5,pyramid,semanticcodebert}. 
Previous studies have demonstrated that explicitly incorporating structural knowledge into models enhances their perceptual understanding of code, thereby aiding software mining tasks~\cite{atom,astt5}, including code translation~\cite{GraphCodeBERT, tree2tree}. 
For example, modeling structural knowledge from AST~\cite{treebert,code2vec,jiang2007deckard,jiang2018shaping}, control flow graphs~(CFG)~\cite{sgattention,semanticcodebert}, and data flow graphs~(DFG)~\cite{GraphCodeBERT}.
However, these syntax-aware methods rely on meticulously crafted model architectures and loss functions that are tailored through training.

Recently, LLMs, scaled to hundreds of billions of parameters, have exhibited remarkable performance across various natural language processing tasks~\cite{openai2023chatgpt,dubey2024llama,anthropic2024introducing}, such as 
GPT~\cite{tang2023chatgpt,gpt4}, 
LLaMA~\cite{touvron2023llama}, 
Qwen~\cite{yang2024qwen2}, etc. 
%
\textcolor{dgreen}{
The LLM-powered code translation has achieved significant progress in recent years~\cite{macedo2024exploring,yang2024exploring,jana2024cotran,luo2024bridging}.
For example, Yang et al.~\cite{yang2024exploring} proposed UniTrans, which introduces auto-generated test cases throughout the whole code translation framework via
translation augmentation and repair. Although UniTrans injects test information and achieves certain effectiveness, it overlooks the role of intrinsic code structure information in assisting the model to understand code semantics.
In fact, mainstream methods based on LLMs often overlook code structural knowledge, whose effectiveness has been validated in prior studies~\cite{semanticcodebert,pyramid,sgattention}. }
Directly combining the carefully designed model architectures and loss functions of classic syntax-aware methods with LLMs is both time- and resource-intensive, and may even be prohibited for proprietary models.
Therefore, research on how to post-incorporate code structural knowledge into off-the-shelf LLMs is limited but crucial.

\subsection{In-context Learning}
Recent studies have shown that LLMs exhibit strong ICL abilities as model and data sizes scale up~\cite{chowdhery2022palm, liu2024incontext, wei2022emergent}. 
The essence of ICL lies in learning by analogy, requiring a prompt context composed of a few demonstration exemplars. 
By concatenating these exemplars with a query, LLMs can make better predictions without updating parameters,
differentiating ICL from traditional model tuning. 
Recently, numerous studies have attempted to leverage ICL for complex tasks, including 
natural language understanding~\cite{liu2021makes},
reasoning~\cite{cot, Lu_Dynamic_ICLR, scarlatos2023reticl},
text generation~\cite{agrawal2022context, shi-etal-2022-xricl, rubin-etal-2022-learning}.
Additionally, several studies aim to provide theoretical justifications and insights into how LLMs learn from a few in-context demonstration exemplars~\cite{von2023transformers, garg2022can}.
Despite significant progress, ICL still faces challenges and requires further exploration. 
such as its sensitivity to the selection of demonstration exemplars~\cite{liu2024lets}.

Various objectives for ICL exemplar selection and tailoring have been explored~\cite{li-etal-2023-unified, cheng-etal-2023-uprise, luo-etal-2021-weakly, luo2024context}.
Mostly ICL exemplar selection typically relies on point-wise similarity~\cite{liu2021makes,gao2022transform,luo-etal-2021-weakly, jiang2007deckard, jiang2018shaping}, 
where a matching score is computed between each candidate exemplar and the input query, followed by the independent selection of the top-$k$ exemplars to construct the context~\cite{luo2024context}.
Although these methods are straightforward and fast, they fail to consider compositional relationships among the selected $k$ exemplars, often leading to redundancy caused by homogeneity. 
Some studies aim to ensure diversity by clustering exemplars into $k$ groups, then selecting the most similar exemplars from each group for a given query~\cite{li2024long,li2022self,li2023mot}.
Besides, some works~\cite{ye2023compositional,levy2022diverse} introduce an auxiliary loss to regularize diversity. 
However, additional diversity loss may hinder the optimization of relevance, and accounting for compositional 
relationships in diversity remains challenging.
Some methods attempt to combine weighted similarity and diversity, but adaptively balancing these two objectives is difficult.
Moreover, these methods do not focus on code translation and are typically syntax-unaware, neglecting the critical syntactic information inherent in source code.
Additionally, \citet{levy2022diverse} attempts to account for local structure coverage in ICL. However, it only considers the structure between a parent node and its child nodes, and the algorithm is rough, needing to work together with complex similarity and diversity regularization losses. 
In contrast, our method is grounded in information theory, providing greater precision and generality. We consider a more comprehensive subtree structure, and our algorithm can be applied directly, guaranteed by submodular theory, without requiring additional losses. 

\section{Conclusion and Future Work}
Code translation aids code migration between languages, boosting productivity in software development.
This work focuses on post-incorporating code structural knowledge into pre-trained models via ICL for code translation.
We revisit exemplar selection in ICL from an information-theoretic perspective, showing that maximizing information coverage provides a more precise and general solution than traditional methods.
We introduce CAST, a surrogate measure based on the maximum subtree coverage ratio of AST.
We show that NP-hard CAST maximization is a submodular maximization problem and propose a greedy algorithm with a $(1-1/e)$-approximate solution of polynomial time complexity.
%
Our method is the first training-free, model-agnostic approach to integrate code structural knowledge into existing LLMs at test time, and both theory and experiments prove their effectiveness and efficiency.
%
This work provides two key insights: 1) Code structural knowledge can be post-incorporated into pre-trained LLMs during inference despite training neglect. 2) Scaling model size or data doesn't lead to code structural knowledge, highlighting the need to consider code syntax in LLMs.

Although our method shows promise in incorporating code structural knowledge into pre-trained models,
several areas for future improvement remain. 
For instance, considering more advanced code structure beyond AST, e.g., CFG and DFG, could provide richer information for software mining in pretrained models.
Besides, introducing a feedback loop during inference, where the model refines exemplar selection based on previous outputs, could further enhance performance. 
%
%

\section*{Acknowledgments}
This research is supported by the National Natural Science Foundation of China~(NSFC) under grant number 62076121, the Major Program~(JD) of Hubei Province~(2023BAA024).

\bibliographystyle{ieeetr}
\bibliography{clean_ref}



\begin{IEEEbiography}[{\includegraphics[width=1in,height=1.25in,clip,keepaspectratio]{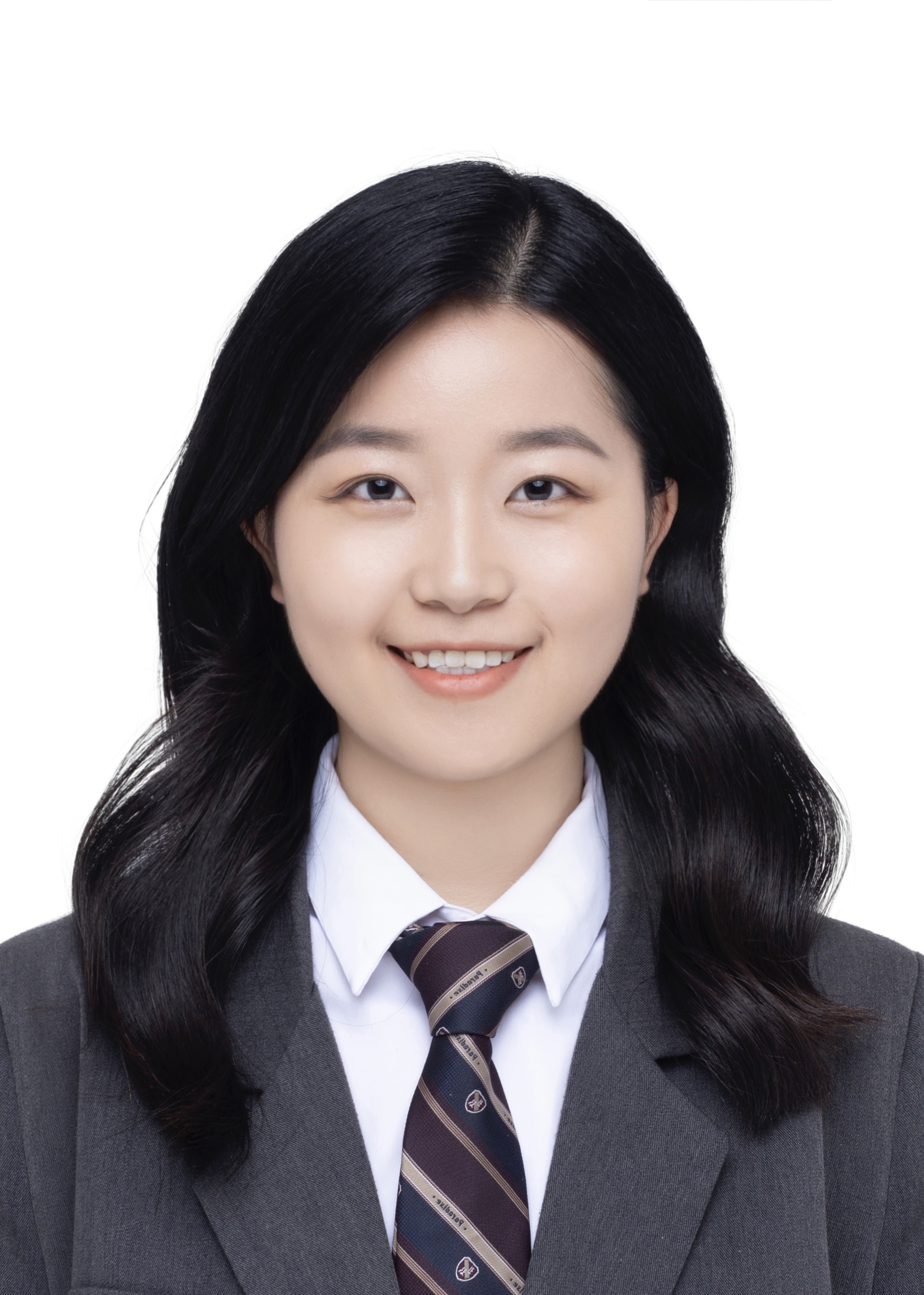}}]{Yali Du}
	received the BEng degree from the School of Computer Science and Technology at Shandong University, China, in 2022. 
	She is currently a PhD student at the School of Artificial Intelligence at Nanjing University and a member of the LAMDA Group. Her research interests include machine learning and data mining, especially on software mining and code representation learning.
\end{IEEEbiography}


\begin{IEEEbiography}[{\includegraphics[width=1in,height=1.25in,clip,keepaspectratio]{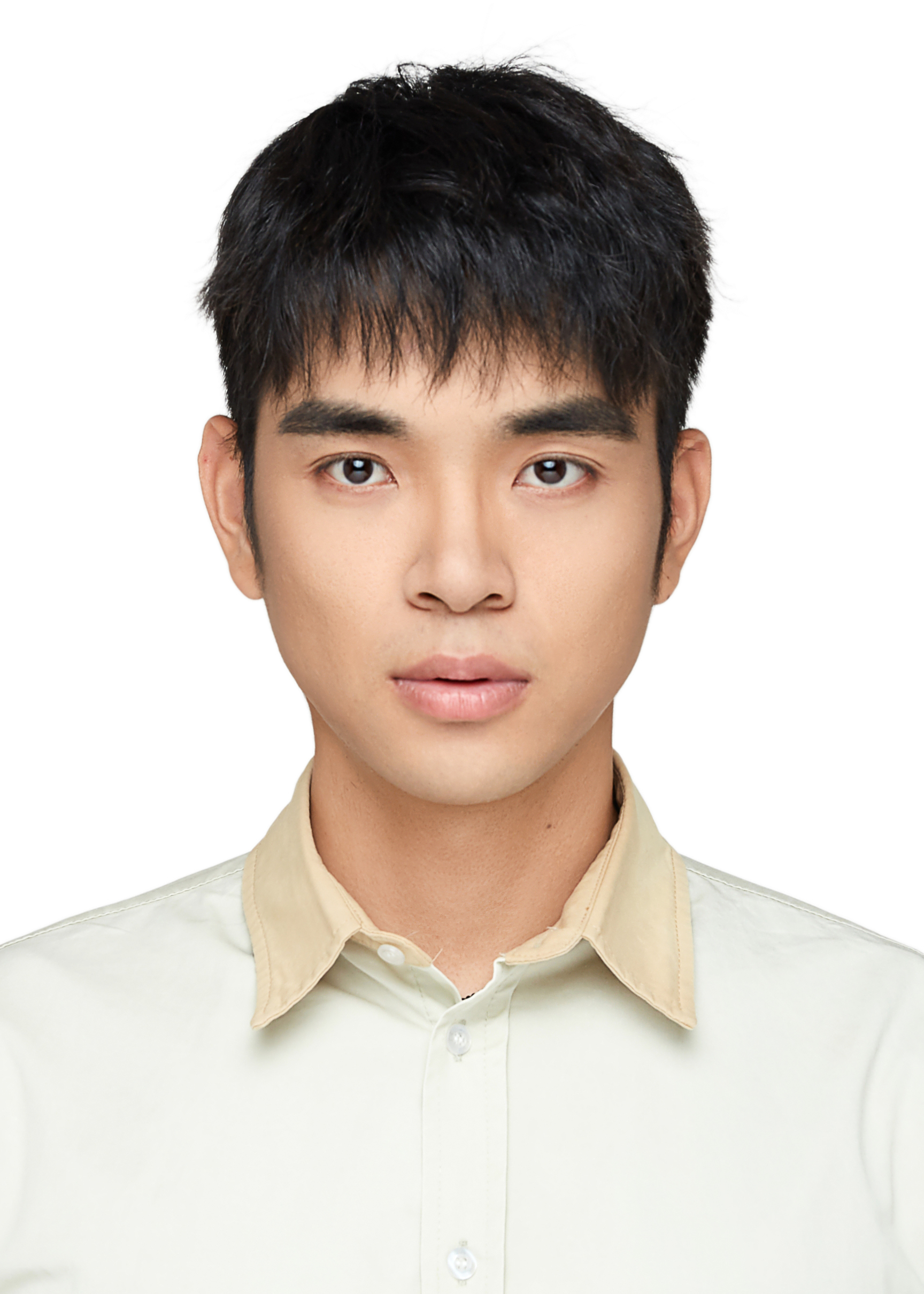}}]{Hui Sun}
	received the BEng degree from the College of Software Engineering at Jilin University, China, in 2019, and the MS degree from the School of Artificial Intelligence at Nanjing University, China, in 2022. 
	He is currently a PhD student in the School of Artificial Intelligence at Nanjing University and is a member of the LAMDA Group. His research interests include transfer learning, semi-supervised learning, and multimodal large language model.
\end{IEEEbiography}


\begin{IEEEbiography}[{\includegraphics[width=1in,height=1.25in,clip,keepaspectratio]{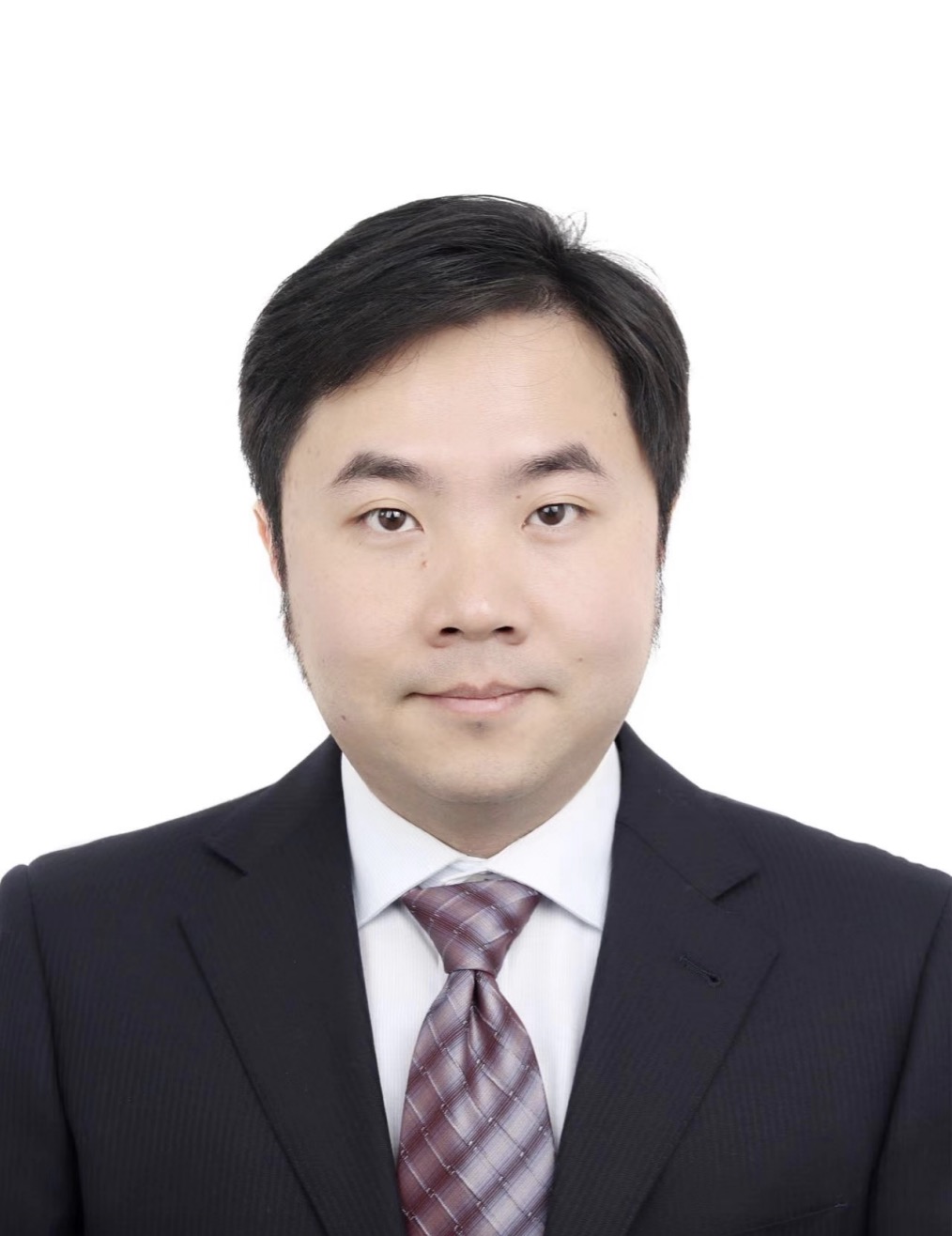}}]{Ming Li}
  (Member, IEEE) is currently a professor with the LAMDA group, the National Key Laboratory for Novel Software Technology, Nanjing University. His major research interests include machine learning and data mining, especially on software mining. He has served as the area chair of IJCAI, IEEE ICDM, etc, senior PC member of the premium conferences in artificial intelligence such as AAAI, and PC members for other premium conferences such as KDD, NeurIPS, ICML, etc. He is the founding chair of the International Workshop on Software Mining. He has been granted various awards including the PAKDD Early Career Award, the NSFC Excellent Youth Award, the New Century Excellent Talents program of the Education Ministry of China, etc.
\end{IEEEbiography}

\vfill

\end{document}